\documentclass[twoside]{article}
\usepackage{graphicx}
\usepackage[T2A]{fontenc}
\usepackage[cp866nav]{inputenc}
\usepackage{amssymb}
\usepackage{amsmath}
\usepackage[eqsecnum]{cmpj}

\title[Introduction to renormalization]%
{Introduction to renormalization}

\author[Yu.Holovatch]
{Yu.Holovatch\refaddr{label1,label2,label3} }
\addresses{
 \addr{label1} Institute for Condensed Matter Physics
of the National Academy of Sciences of Ukraine,\\
1 Svientsitskii Str., 79011 Lviv, Ukraine
 \addr{label2} Institute f\"ur Theoretische Physik,
Johannes Kepler Universit\"at Linz, A--4040, Linz, Austria
 \addr{label3} Ivan Franko National University of Lviv,
79005 Lviv, Ukraine}

\begin{document}

\maketitle

\begin{abstract}
In these lectures I discuss peculiarities of the critical
behaviour of ``non-ideal'' systems as it is explained by the
renormalization group approach. Examples considered here include
account of the single-ion anisotropy, structural disorder,
frustrations. I introduce main ideas of renormalization and show
how  it serves the explanation of typical features of criticality
in the above systems: softening of the phase transition, changes
in the universality class, complicated effective critical
behaviour.
\pacs 05.50.+q, 05.70.Jk, 64.60.Ak
\end{abstract}

{\small \tableofcontents}

\section{Introduction} \label{I}

There exist many ways one can choose to approach the subject of
these lectures. Taken that the presentation is limited in time
(spent by the students and lecturers at the Caribbean seashore
during the Mochima school in theoretical physics) and in space
(given by the Editors of this volume) this choice becomes a
difficult one. The way of presentation I decided to follow was
chosen for several reasons. Extremely high theoretical level of
certain modern renormalization group (RG) studies of criticality
in different systems (i.e. the ``language'' of these studies)
sometimes does not allow uninitiated reader to follow the
derivations and even to understand the problem statement and/or
physical consequences of the results. Therefore the goal would be
to provide a minimal vocabulary, explaining main notions as simple
as possible. On the other hand, it is tempting to use such a
simple vocabulary to make a short review of state-of-the art RG
studies in a certain domain. RG explanation of criticality in
``non-ideal'' 3d systems might be a good candidate for such a
domain: it is a subject of ongoing activity where important
results have recently been obtained and still a lot is to be done.

The following account will serve  this purpose: after mentioning
several examples of criticality and scaling in condensed matter
physics (section~\ref{II}) and introducing model Hamiltonians of
the ``non-ideal'' systems we shall be interested in
(section~\ref{III}) I shall give the main ideas and notions of the
renormalization taking as an example a simple 1d Ising model
(section~\ref{IV}). Once the reader is acquainted with the  RG
transformation, its flow and fixed points, stability, universality
and scaling, I shall pass to the ``non-ideal'' systems showing how
to obtain their effective Hamiltonians and to the reviewing of
recent results in this domain (sections~\ref{V} and \ref{VI}).
Some conclusions and outlook are given in section~\ref{VII}.

\section{Criticality and scaling} \label{II}

\begin{figure}[h]
\begin{picture}(80,170)
\put(20,15){\parbox[t]{110mm}{\includegraphics[width=
60mm,angle=0]{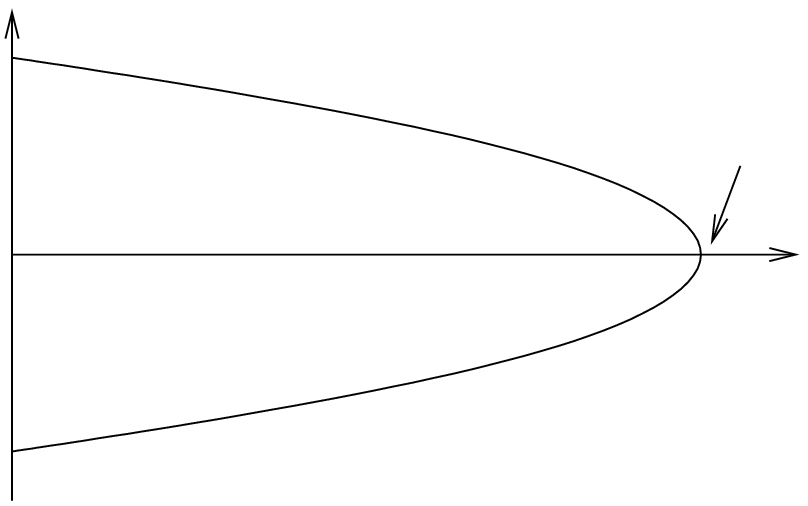}  }}
 \put(10,115){\parbox[t]{10mm}{$\rho$}}
\put(170,95){\parbox[t]{10mm}{$T_{\mathrm{c}}$}}
     \put(180,55){\parbox[t]{10mm}{$T$}}
     \put(90,105){\parbox[t]{20mm}{\bf \em liquid}}
     \put(90,25){\parbox[t]{20mm}{\bf \em gas}}
  \put(100,00){\parbox[t]{10mm}{(a)}}
 \put(220,15){\parbox[t]{110mm}{\includegraphics[width=
60mm,angle=0]{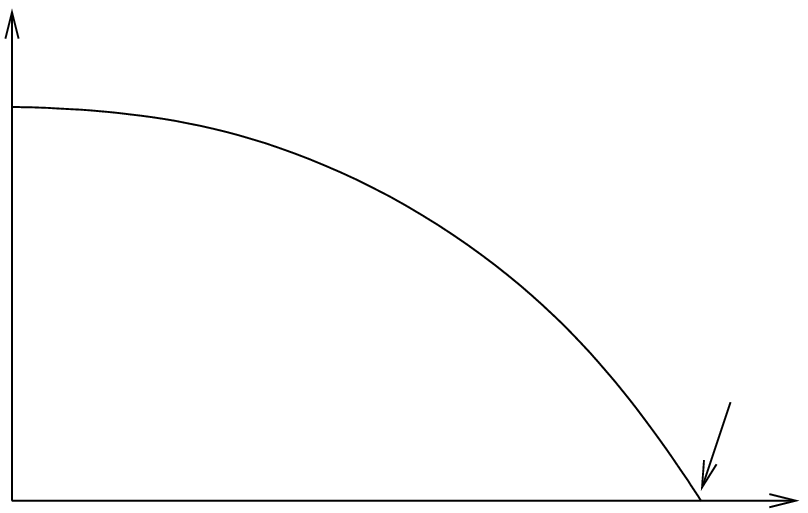} }}
 \put(200,115){\parbox[t]{10mm}{$M$}}
 \put(370,45){\parbox[t]{10mm}{$T_{\mathrm{c}}$}}
    \put(380,05){\parbox[t]{10mm}{$T$}}
   \put(305,00){\parbox[t]{10mm}{(b)}}
 \end{picture}
\caption{\label{fig1}
 Order parameters for the critical point of
fluids (a) and Curie point of magnets (b).}
\end{figure}

It is generally recognized that the term ``critical point'' was
introduced in 1869 by Tho\-mas An\-drews who studied a special
point for carbon dioxide at about 31$^{\circ}$C and 73 atmospheres
pressure where the properties of liquid and of gas become
indistinguishable \cite{history}. Approaching the critical point
$T_{\mathrm{c}}$ from below, the liquid-gas density difference
obeys a power law {\em scaling} governed by the {\em critical
exponent} $\beta$:
\begin{equation}\label{2.1}
\rho_{\rm L} - \rho_{\rm G} \propto (T_{\mathrm{c}} - T)^{\beta},
\qquad T \rightarrow T_{\mathrm{c}}^-,
\end{equation}
as shown in the figure~\ref{fig1}a.
Similar power laws were found for some other physical quantities,
describing their approach to zero or singularity at
$T_{\mathrm{c}}$. Less known is the fact, that the mean field
values of critical exponents (which are integers or their simple
relations, e.g. $\beta=1/2$ for equation~(\ref{2.1})) were
questioned already over a century ago
\cite{Verschaffelt00}\footnote{I thank Reinhard Folk for pointing
me the work of Jules-\'Emile Verschaffelt. }. Therefore much time
passed before there appeared the theory that both explained
scaling behaviour (\ref{2.1}) found in condensed matter physics,
and, on a larger scale, not only in physics, as well as offered
quantitatively accurate predictions for the exponents. As we now
know this theory is based on the renormalization group formalism
\cite{Wilson} and has its origin in quantum field problems
\cite{RGbooks}.

Next experimental observations of critical behaviour arrived in
1895, when Pierre Curie showed that the ferromagnet iron also
displayed a special point. This point is the highest temperature
at which iron can be permanently magnetized at zero external
magnetic field which is now called the Curie point. Curie himself
noticed the parallelism between density-temperature curve at
constant pressure for carbon dioxide and magnetization-temperature
curves at constant magnetic field for iron. Behaviour of the {\em
order parameter}, spontaneous magnetization $M$, in the vicinity
of Curie point (we denote it $T_{\mathrm{c}}$ as well) is governed
by the familiar scaling law, cf. equation~(\ref{2.1}):
\begin{equation}\label{2.2}
M \propto (T_{\mathrm{c}} - T)^{\beta}, \qquad T \rightarrow
T_{\mathrm{c}}^-.
\end{equation}
It appears that different magnetic systems possess different sets
of critical exponents but there exist wide classes of magnets
characterized by the same values of exponents. For example,
uniaxial magnets (like ${\rm FeF_2}$, ${\rm MnF_2}$, ${\rm
K_2CoF_4}$, ${\rm Rb_2CoF_4}$) possess the same critical
exponents. Moreover, they coincide with those of simple fluids
(${\rm Xe}$, ${\rm SF_6}$, ${\rm CO_2}$ \dots) and some other
systems. In such a case it is said that these systems belong to
the same {\em universality class}. The reasons for this as well as
the main characteristics of a system which define the universality
classes will be cleared up in the subsequent sections.

As far as the second derivatives of Gibbs free energy (isothermal
magnetic susceptibility for magnets and isothermal compressibility
for fluids) diverge at $T_{\mathrm{c}}$ such {\rm phase
transitions} are referred  to as the second order phase
transitions. This classification is due to Paul Ehrenfest. The
order parameter does not possess a discontinuity at
$T_{\mathrm{c}}$: it is an example of a continuous phase
transition. Phenomena which occur in the vicinity of the 2nd order
phase transition point are called critical phenomena. Changes of
system structure at $T_{\mathrm{c}}$ are reflected in the
behaviour of the pair correlation function $G_2(r)$
(density-density function for fluids and
magnetization-magnetization one for magnets) and described by two
universal exponents $\eta$ and $\nu$. The exponential decay of
$G_2(r)$ with distance at $T>T_{\mathrm{c}}$ transforms to the
power law one:
\begin{equation}\label{2.3}
G_2(r) \sim \re^{-r/\xi}\,, \qquad T>T_{\mathrm{c}}\,;  \qquad
G_2(r)\sim \frac{1}{r^{d-2+\eta}}\,,  \qquad T=T_{\mathrm{c}}\,,
\end{equation}
 where $\xi$ is a correlation length which diverges at $T_{\mathrm{c}}$:
 $\xi\sim |T-T_{\mathrm{c}}|^{-\nu}$, $T \rightarrow T_{\mathrm{c}}$.

The XX century witnessed numerous experiments performed after
pioneering works of Thomas Andrews and Pierre Curie where it was
shown that criticality and scaling accompany not only the second
order thermodynamic phase transitions. They are found in quantum
phase transitions, percolation, non-equilibrium dissipative phase
transitions. Properties  of long flexible polymer chains in good
solvents are described in terms of critical phenomena as well.
This list can be continued. However already the above mentioned
field of phenomena is very wide, especially if one takes into
account that any of them occurs in objects differing by their
microscopic nature. Thermodynamic 2nd order phase transitions
occur in magnets, ferroelectrics, quantum liquids; percolation
phenomena occurs in resistor networks, random magnets, gas
filters. Nevertheless there are at least two essential features
common for the above mentioned phenomena. These are: {\it the
singular character of change of properties} in certain critical
points and {\it universal behaviour}  in the vicinity of these
points. As it became clear now the reason for both is anomalous
growth of fluctuations and their correlation at very high
distances in the vicinity of critical points. Correlation range
becomes the only characteristic scale of the system and this
causes the insensitivity of its behaviour to the so-called
microscopic parameters. In particular this results in the {\em
scale invariance} of the system where critical phenomena occur.

Since the pioneering work of Ising \cite{Ising25}, theoretical
description of many-particle systems is often based on the
so-called classical spin models. They appeared to be of primary
importance to reveal the main features of criticality. In the next
section \ref{III} we shall discuss several models currently used
in describing the critical phenomena, the analysis of which will
be a subject of subsequent sections.

\section{Non-ideal spin Hamiltonians: single ion anisotropy,
structural disorder, frustrations} \label{III}

In classical spin models, each particle is imitated as a vector
located on a lattice site and ``interacting'' with other vectors.
The interaction is chosen to mimic the interparticle one whereas
the dimensionality of the vector is equal to the number of
internal degrees of freedom of the particle. Below we start with
what we call an ideal system, which is a $m$-vector model and show
how to include different types of non-idealities within a spin
model formalism.

\subsection{An ideal system: $m$-vector model}\label{IIIa}
The model describes a system of $m$-dimensional classical vectors
(``spins'') located in the sites of $d$-dimensional hypercubic
lattice. The Hamiltonian of the $m$-vector model in the absence of
an external magnetic field reads:
\begin{equation}\label{3.1}
{\cal H} =  - \frac{1}{2}\sum_{\bf R, R'} J\left(|{\bf R} - {\bf
R'}|\right) \vec{S}_{{\bf R}} \vec{S}_{{\bf R'}}\,,
\end{equation}
where $J(|{\bf R} - {\bf R'}|)$ is an interaction between spins
$\vec{S}_{{\bf R}}$ and $\vec{S}_{{\bf R'}}$ located in sites
${\bf R}$ and ${\bf R'}$ (we shall consider it to be the
short-range one and of ferromagnetic origin: $J(R)>0$) and
$\vec{S}_{{\bf R}} \vec{S}_{{\bf R'}}$ means a scalar product.
Such a model appears as a natural generalization of the Ising
model for the case of $m$-component spin \cite{Stanley68} and
serves as a basic model in describing phase transitions in systems
with multicomponent order parameter. Below, we shall be mainly
interested in the 3d systems. However, for the sake of
completeness let us recall that ferromagnetic ordering does not
occur in this model at $d=1$ \cite{Stanley68}, it does not occur
for $d=2$ and $m>1$ either \cite{continuous}. Whereas the
celebrated Onsager solution \cite{Onsager44} of the 2d Ising model
($d=2$, $m=1$) brings about the second order phase transition with
non-trivial values of the critical exponents: $\beta=1/8$,
$\nu=1$, $\eta=1/4$, $\gamma=7/4$, $\alpha=0$ (exponents $\gamma$
and $\alpha$ govern power law scaling of the magnetic
susceptibility and specific heat, the latter being logarithmically
divergent for the 2d Ising model). At $d=3$, ferromagnetic
ordering occurs at any $m$ (see table \ref{tab1} for typical
values of exponents), and $d=4$ is the {\em upper critical
dimension} of the problem: exponents attain their mean field
values for $d\geqslant  4$.
\begin{table}[ht]
\caption{The standard values of the critical exponents of the 3d
$m$-vector model (obtained in~\protect\cite{Guida98} from the
high-order renormalization group expansions). \label{tab1} }
\begin{center}
{\begin{tabular}{|c|c|c|c|c|c|} \hline $m$ & $\gamma$ & $\nu$ &
$\eta$ & $\beta$ & $\alpha$ \\ \hline 1 & $1.2396(13)$ &
$0.6304(13)$ & $0.0335(250)$ & $0.3258(14)$ & $0.109(4)$ \\ \hline
2 & $1.3169(20)$ &
$0.6703(15)$ & $0.0354(250)$ & $0.3470(16)$ & $-0.011(4)$ \\
\hline 3 & $1.3895(50)$ & $0.7073(35)$ & $0.0355(250)$ &
$0.3662(25)$ & $-0.122(10)$ \\ \hline
\end{tabular}}
\end{center}
\end{table}

We refer to the above model as to the ideal one: it describes the
lattice system without any defects of structure or complexities of
interaction which are often encountered in real systems. Let us
show how those can be considered within the same formalism.

\subsection{Single-ion anisotropy}\label{IIIb}

Real substances often are anisotropic. For instance, in cubic
crystals one expects the spin interaction to react on the lattice
structure (crystalline anisotropy) suggesting additional terms in
the Hamiltonian, invariant under the cubic group \cite{Aharony76}.
Such a single-ion anisotropy breaks the rotational symmetry of the
$m$-vector magnet (\ref{3.1}) and the Hamiltonian reads:
\begin{equation}\label{3.2}
{\cal H} =  - \frac{1}{2}\sum_{\bf R, R'} J\left(|{\bf R} - {\bf
R'}|\right) \vec{S}_{{\bf R}} \vec{S}_{{\bf R'}} + V \sum_{\bf R}
\sum_{i=1}^m \left(S^i_{\bf R}\right)^4,
\end{equation}
where $V$ defines anisotropy strength and makes the order
parameter to point either along the edges ($V>0$) or along
diagonals ($V<0$) of a  $m$-dimensional hypercube. Therefore a
model with the spin Hamiltonian (\ref{3.2}) is often called a {\em
cubic model}.

An interesting phenomenon is observed in 3d cubic
magnets: for low values of $m$ their critical exponents coincide
with those of the ``ideal'' $m$-vector model (one speaks about
isotropization of critical fluctuations), whereas for large $m$
they belong to the new  universality class. The value
$m_{\mathrm{c}}^{\rm cub}$ at which one regime is changed to the
other one is called the marginal dimension. Moreover, transition
to the low-temperature phase may also occur via the first-order
scenario.

\subsection{Structural disorder}\label{IIIc}

To treat structural disorder within the lattice model one usually
introduces random variables into the Hamiltonian (\ref{3.1}) and
couples them to spin degrees of freedom. The new model can either
mimic a quenched system (new variables are randomly distributed
and fixed in a certain configuration) or an annealed, equilibrium
one \cite{Brout59}. Here, we shall be interested in changes of
critical behaviour caused by quenched disorder. Furthermore, we
shall consider two different examples, showing how to introduce
disorder via dilution and random anisotropy\footnote{Another
option would be to consider a random field disorder
\cite{Belanger91}. However, we do not introduce it since an
appropriate model differs from those considered here by its upper
critical dimension and this will make our account even broader.
The same concerns strong dilution at the percolation threshold.}.

\subsubsection{Random-site dilution}\label{IIIc1}
To describe dilution, one may introduce the random-site $m$-vector
model, considering situation, when spins $\vec{S}_{{\bf R}}$ in
(\ref{3.1}) occupy only a part of the lattice sites, $N_1$, the
rest $N-N_1$ sites being empty (or occupied by non-magnetic
atoms). Magnetic and non-magnetic sites are randomly distributed
and fixed in a certain configuration. The model Hamiltonian reads:
\begin{equation} \label{3.3}
{\cal H} =  - \frac{1}{2}\sum_{\bf R, R'} J\left(|{\bf R} - {\bf
R'}|\right) c_{{\bf R}}c_{{\bf R'}}\vec{S}_{{\bf R}} \vec{S}_{{\bf
R'}}\,,
\end{equation}
where $c_{{\bf R}}$ are the occupation numbers:
\begin{equation} \label{3.4}
c_{{\bf R}} = \left \{ \begin{array}{ll} 1, & \quad \mbox{site
${\bf R}$ is occupied},
\\
0, & \quad \mbox{site ${\bf R}$ is empty}.
\end{array}
\right.
\end{equation}
Here we do not touch upon the phenomena occuring near the
percolation threshold $c_{\rm perc}$ and consider the so-called
weak dilution, $c=N_1/N\gg c_{\rm perc}$. To complete the model
one should choose a certain distribution function for the
occupation numbers $c_{{\bf R}}$. Let us consider the case when
the site is occupied with the probability $c$ and is empty with
the probability $(1-c)$ and this probability does not depend on
the occupation numbers on the neighbouring sites. Such a situation
corresponds to the following occupation probability ${\cal P}
(\{c_{{\bf R}} \})$:
\begin{equation} \label{3.5}
{\cal P} (\{c_{{\bf R}} \})=\prod_{\bf R} p(c_{{\bf R}}),
\hspace{1cm} p(c_{{\bf R}})= c \delta (c_{{\bf R}}-1) + (1-c)
\delta (c_{{\bf R}}).
\end{equation}
For different $m$, the above model describes magnetic phase
transitions in crystalline alloys of uniaxial magnets and their
non-magnetic isomorphs ${\rm Fe_xZn_{1-x}F_2}$, ${\rm
Mn_xZn_{1-x}F_2}$ \cite{Folk03}, diluted Heisenberg-like magnets
\cite{Pelissetto02,Dudka03}, superfluid phase transition in He$^4$
in porous medium \cite{Yoon97}. Similar to the cubic anisotropy
systems, there exists a marginal dimensionality
$m_{\mathrm{c}}^{\rm dil}$ separating two different scenarios: for
$m<m_{\mathrm{c}}^{\rm dil}$ dilution causes a change in the
universality class, whereas the asymptotic critical exponents of
$m>m_{\mathrm{c}}^{\rm dil}$-component systems remain unchanged
under dilution. The so-called Harris criterion allows us to
determine this marginal dimension from the pure system heat
capacity behaviour \cite{Harris74}. It states that critical
exponents are not altered by dilution, if the heat capacity of the
pure system does not diverge ($\alpha<0$). From Table \ref{tab1}
one concludes that in three dimensions it is the Ising model which
changes the universality class.

\subsubsection{Random anisotropy}
Another way of introducing randomness to the model (\ref{3.1}) is
to consider the case, when each spin is subjected to a local
anisotropy of random orientation. The resulting Hamiltonian reads
\cite{Harris73}:
\begin{equation}\label{3.6}
{\cal H} =  - \frac{1}{2}\sum_{\bf R, R'} J(|{\bf R} - {\bf R'}|)
\vec{S}_{{\bf R}} \vec{S}_{{\bf R'}}
-D\sum_{{\bf R}} (\hat {x}_{\bf
R}\vec{S}_{\bf R})^{2}.
\end{equation}
Here,  $D>0$ is the strength of the anisotropy and $\hat {x}_{\bf
R}$ is a random unit vector pointing in the direction of the local
anisotropy axis. The random anisotropy model (\ref{3.6}) is
relevant to the description of a wide class of disordered magnets.
It was first introduced to describe magnetic properties of
amorphous alloys of rare-earth compounds with aspherical electron
distributions and transition metals \cite{Harris73}. Today the
majority of the amorphous alloys containing rare-earth elements
are recognized to be random anisotropy magnets \cite{ramreviews}.

As in the random-site case, the model should be completed by
choosing a certain distribution $p(\hat x_{\bf R})$ for the random
variables $\hat x_{\bf R}$. Most often, two different
distributions are considered \cite{Aharony75}. The first is an
isotropic one, where the random vector $\hat x_{\bf R}$ points
with equal probability in any direction in the $m$-dimensional
hyperspace:
\begin{equation}\label{3.7}
p(\hat x_{\bf R}) \equiv\left(\int \rd^m \hat x_{\bf
R}\right)^{-1}= \frac{\Gamma(m/2)}{2\pi^{m/2}}\,.
\end{equation}
Here $\Gamma(x)$ is Euler gamma-function, and the right-hand side
presents the volume of the $m$-dimensional hypersphere of unit
radius. This distribution mimics an amorphous system without any
preferred direction. The second distribution restricts the vector
$\hat x_{\bf R}$ to point with equal probability along one of the
$2m$ directions of the axes $\hat k_i$ of a hypercubic lattice:
\begin{equation}\label{3.8}
p(\hat x_{\bf R}) = \frac{1}{2m}\sum_{i=1}^m\left[\delta^{(m)}
(\hat x_{\bf R}-\hat k_i)+\delta^{(m)}(\hat x_{\bf R}+\hat k_i)\right],
\end{equation}
where $\delta(y)$ are Dirac $\delta$-functions. This distribution
(sometimes called a cubic one) corresponds to a situation when an
amorphous magnet still remembers the initial cubic lattice
structure.

It is generally believed that ferromagnetism does not exist for
the 3d model (\ref{3.6}) with the isotropic random axis
distribution (\ref{3.7}) \cite{ramreviews}. However, anisotropic
distribution (\ref{3.8}) leads to a magnetically ordered low
temperature phase and the transition belongs to the random site
Ising model universality class \cite{Dudka05}.

\subsection{Frustrations}\label{IIId}

An archetype of a model describing the effect of frustrations is
the model of stacked triangular antiferromagnet \cite{Kawamura88}.
In this model, the antiferromagnetically interacting spins are
placed on the sites of 2d triangular lattices stacked in register
along the orthogonal direction. The sign of the interlayer
interaction is unimportant, because there are no frustrations in
orthogonal direction. The Hamiltonian reads:
\begin{equation} \label{3.9}
{\cal H} =  - \frac{1}{2}\sum_{\langle {\bf R,R'}\rangle} J \,
\vec{S}_{\bf R} \vec{S}_{\bf R'},\qquad
  J = \left \{ \begin{array}{ll} J_1 < 0, &  \mbox{inside a
plane,}
\\
J_2\,, & \mbox{between planes.}
\end{array}
\right.
\end{equation}
Sum in (\ref{3.9}) runs over the nearest neighbours of the above
described lattice. Systems which are characterized by the
Hamiltonian (\ref{3.9}) exhibit noncollinear spin ordering. An
example is given by the famous 120$^\circ$ structure: each spin in
a layer forms 120$^\circ$ angles with the neighbouring spins.
Although  model (\ref{3.9}) is formulated for general $m$, of most
interest are values $m=2$ and $m=3$. Namely for these values of
$m$ the model has experimental realizations
\cite{Pelissetto02,Delamotte04} and describes noncollinear
ordering of stacked triangular antiferromagnets as ${\rm VCl_2}$,
${\rm VBr_2}$, ${\rm CsMnBr_3}$,  and helical magnets as ${\rm
Ho}$, ${\rm Dy}$, $\beta-{\rm MnO_2}$. In the latter substances,
frustration is induced by the competition of ferromagnetic
nearest-neighbour  and antiferromagnetic next-nearest-neighbour
interactions, which acts only along one lattice axis: $J_1>0$ and
$J_2<0$, correspondingly. When the ratio $J_1/J_2$ exceeds a
critical value, in the low-temperature phase the spins align
ferromagnetically in a plane and form a spiral along the
orthogonal axis. Moreover, at $m=3$ Hamiltonian (\ref{3.9}) also
describes A/B transition in He$^3$.

Numerous experimental and MC studies performed so far have not
lead to the definitive conclusion about the order of transition
into non-collinear state. There is no unique answer from the
theoretical viewpoint either \cite{Pelissetto02,Delamotte04}.

\section{Renormalization} \label{IV}

In order to explain the main ideas of the renormalization group
(RG) theory and to show how this method works in practice we shall
study the critical behaviour of the 1d Ising model by means of the
RG approach. The explanations given below are due to Michael
Fisher \cite{Fisher}, who compared this study with the
Bohr-Sommerfeld picture in quantum mechanics. Passing from the
classical mechanics to the full account of quantum mechanics,
Bohr-Sommerfeld's picture represents only a crude approximation.
Nevertheless, it introduces some important ideas, just like RG
study of the 1d Ising model enables one to introduce ideas of
renormalization and scaling in the critical region.

\subsection{RG transformation}\label{IVa}

We start from the Hamiltonian of the 1d Ising model in the
presence of an external magnetic field $H$, which being divided by
$k_{\mathrm{B}}T$ is written in the following form:
\begin{equation}
-{\cal H}_{\rm eff} = K \sum_{j=1}^{N}  S_j S_{j+1} + h
\sum_{j=1}^{N}  S_j + C \sum_{j=1}^{N}  1, \label{12.23}
\end{equation}
where $K=J/(k_{\mathrm{B}}T)$, $h=H/(k_{\mathrm{B}}T)$, and the
last term $C \sum_{j=1}^{N} 1=NC$ is added for the convenience of
forthcoming calculations. We call this temperature-dependent
Hamiltonian an effective one, ${\cal H}_{\rm eff}$. Note, however,
that contrary to the effective Hamiltonians considered  in the
forthcoming section \ref{V}, it contains full information about
the spin Hamiltonian. Given $\{K,h,C\}$ one completely specifies
${\cal H}_{\rm eff}$: thus it can be regarded as a point in a
space of $\, {\rm 3} \,$ parameters $\{K,h,C\}$. With change of
$\{K,h,C\}$, this point moves.

One of the first approaches to the RG is to regard it as a
specific way of calculating the partition function $Z_N\{{\cal
H}\}$:
\begin{equation}
Z_N\{{\cal H}_{\rm eff}\} = \frac{1}{2^N} \sum_{S_1=\pm 1} \dots
\sum_{S_N=\pm 1}  \re^{-{\cal H}_{\rm eff}}, \label{12.24}
\end{equation}
where we have normalized the expression for the partition function
by the partition function of the ideal model ($Z^{{\rm
ideal}}=2^N$) to obtain $Z_N\{{\cal H}_{\rm eff}\}
\rightarrow_{T\rightarrow \infty} 1$. The RG method of calculating
$Z_N$ is to step-by-step perform the summation in (\ref{12.24}) in
such a way as to try as  much as possible to preserve the system
as it used to be prior the summation.

\begin{figure} [ht]
\centerline{\includegraphics[width=8cm] {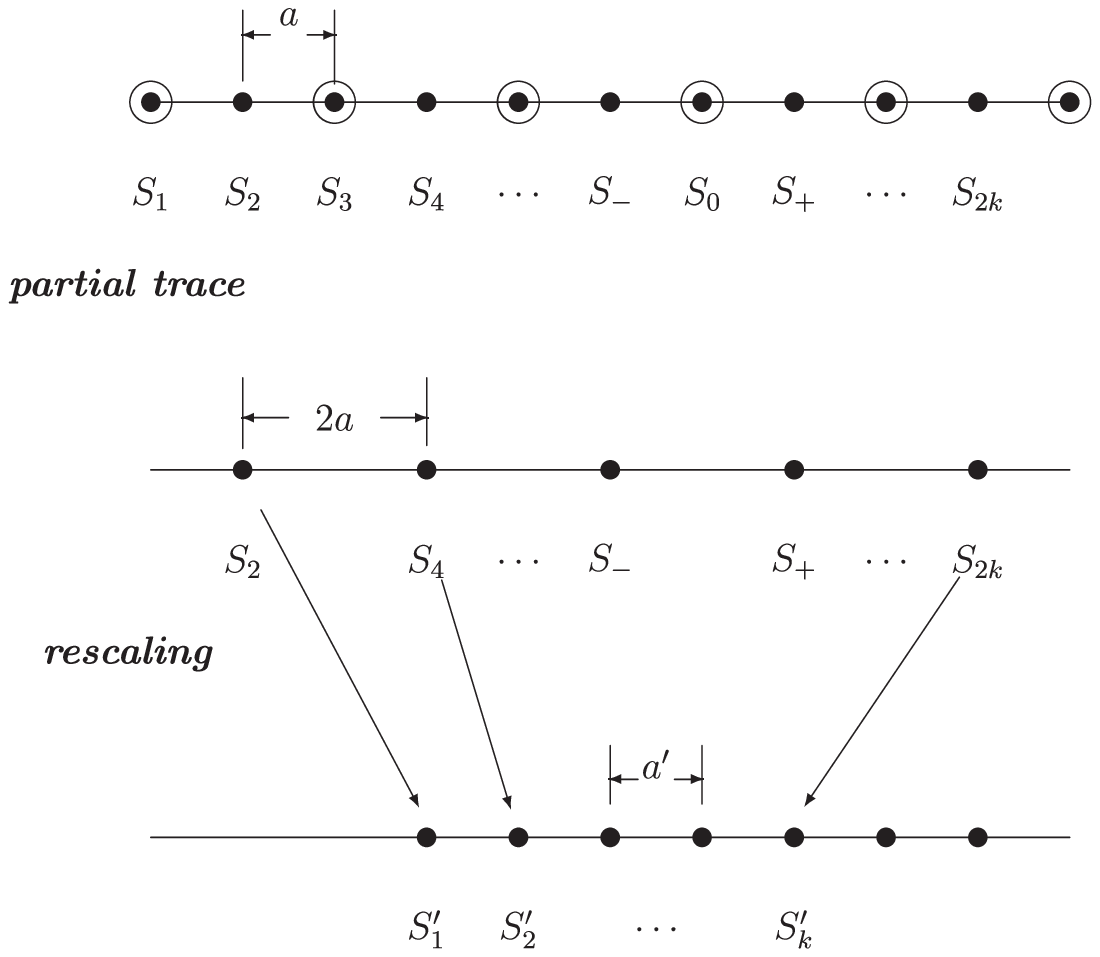}} \caption {{\it
Decimation} procedure followed by the {\it spatial rescaling}.
Summation over every second spin variable is performed and then
all lengths are rescaled in such a way that the new lattice
spacing $a^{\prime}$ is equal to the old one $a$. Now any distance
$R$ in the original lattice becomes $R^{\prime}= {1}/{2} R$ in the
new one measured in the units of the lattice spacing. \label{fig2}
}
\end{figure}

Let us do this by means of {\it decimation} procedure: performing
summation over every second spin variable (see figure~\ref{fig2}).
To show the changes which are introduced by summation over certain
spin variable $S_0$ let us represent the total Boltzmann weight in
the factorized form:
\begin{equation}
\re^{-{\cal H}_{\rm eff}} = \cdots \re^{KS_-S_0 +
\frac{1}{2}h(S_-+S_0)+C} \re^{KS_0S_+ + \frac{1}{2}h(S_0+S_+)+C}
\cdots= \cdots P(S_-,S_0) P(S_0,S_+) \cdots \label{12.26}
\end{equation}
Summation over $S_0$ which enters only $P(S_-,S_0)$ and $P(S_0,S_+)$
will lead to a new function $P^{\prime}(S_-,S_+)$, defined by:
\begin{equation}
P^{\prime}(S_-,S_+) = \frac{1}{2} \sum_{S_0=\pm 1}
P(S_-,S_0) P(S_0,S_+).
\label{12.27}
\end{equation}
Factor ${1}/{2}$ is included in (\ref{12.27}) because with each
spin eliminated one must remove factor $\frac{1}{2}$ from the
normalizing factor in (\ref{12.24}). Now the RG idea is to express
the new factor $P^{\prime}$ in the same form as the initial one:
\begin{equation}
P^{\prime}(S_-,S_+) = \re^{K^{\prime}S_-S_+ +
\frac{1}{2}h^{\prime}(S_-+S_+)+C^{\prime}}. \label{12.28}
\end{equation}
The new parameters define the {\it renormalized Hamiltonian}:
\begin{equation}
{\cal H_{\rm eff}^{\prime}} = {\cal H_{\rm eff}^{\prime}}
\{K^{\prime}, h^{\prime}, C^{\prime} \}, \label{12.29}
\end{equation}
and ${\cal H_{\rm eff}^{\prime}}$ has a half of initial spins. The
result can be formally written as :
\begin{equation}
{\cal H_{\rm eff}^{\prime}} = {\cal R}_b [{\cal H}_{\rm eff} ],
\label{12.30}
\end{equation}
with a  spatial rescaling factor $b$ (in our case $b=2$, see
figure~\ref{fig2}). The resulting number of spins $N^{\prime}$ is
connected with the initial one by $N^{\prime} = N/b$, and in the
case of $d$-dimensional system this is generalized to:
\begin{equation}
N^{\prime} = N/b^d.
\label{12.31}
\end{equation}

Performing summation (\ref{12.27}) and expressing the result in
the renormalized form (\ref{12.28}) one can get expressions for
the renormalized variables $K^{\prime}, h^{\prime}, C^{\prime}$
(it is proposed to the interested reader to do this).

Now the renormalized model is characterized by the Hamiltonian
${\cal H_{\rm eff}^{\prime}}( K^{\prime}, h^{\prime}, C^{\prime})$
having the form  similar to the initial one, though the lattice
constant after taking the sum over every second spin is equal to
$2a$ (see figure~\ref{fig2}). In order to have the renormalized
model looking as the initial one we {\it rescale} all lengths in
such a way that the new lattice spacing $a^{\prime}$ is equal to
the old one. Now any distance $R$ in the original lattice becomes
$R^{\prime}= {1}/{2} R$ in the new one measured in the units of
the lattice spacing. For the arbitrary $b$ we have the following
mapping:
\begin{equation}
R \Rightarrow  R^{\prime} = R/b.
\label{12.35}
\end{equation}
Now let us have a look at the behaviour of the spin-spin
correlation function $\langle S_0 S_R \rangle$. First let us
renumber the remaining spins to have the labels arranged in the
subsequent order (see figure~\ref{fig2}): $ S_2 \Rightarrow
S^{\prime}_1, \quad S_4 \Rightarrow  S^{\prime}_2, \quad \dots,
\quad S_{2k} \Rightarrow  S^{\prime}_k, \dots $ Or, taking into
account (\ref{12.35}), $ S_{2R^{\prime}} =
S_{R^{\prime}}^{\prime}$. Since the remaining spins $S^{\prime}$
have not changed under the renormalization procedure, the
renormalized correlation function is equal to the original one:
\begin{equation}
\langle S_0 S_{2R^{\prime}} \rangle =
\langle S_0^{\prime} S_{R^{\prime}}^{\prime} \rangle.
\label{12.38}
\end{equation}
It follows that if the original correlation length is
$\xi=\xi({\cal H}_{\rm eff})$ then the renormalized correlation
length is two times smaller. Or, for a general $b$:
\begin{equation}
\xi({\cal H}_{\rm eff}) = b \xi({\cal H}_{\rm eff}^{\prime}).
\label{12.39}
\end{equation}
The RG procedure has the effect of shrinking the correlation
length. Recalling that $\xi$ becomes infinite at
$T=T_{\mathrm{c}}$ one can state, that the RG procedure is driving
a system away from criticality (if it was not critical). Already
here it is seen that the RG transformation we are considering has
a deep connection with the critical properties of a system.

\subsection{Fixed points and RG flows}\label{IVb}

\begin{figure} [ht]
\centerline{\includegraphics[width=7cm] {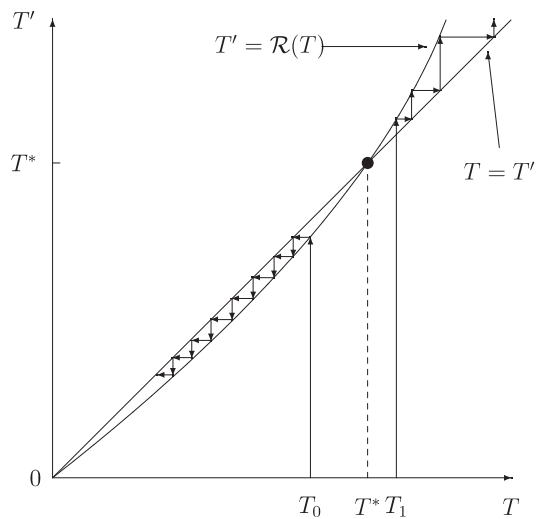}}\caption
{Temperature renormalization function ${\cal R}(T)$. ${\cal R}(T)$
crosses the line $T^{\prime}=T$ at the fixed point $T=T^*$. When
$T$ is smaller ($T=T_0$) or greater ($T=T_1$) than $T^*$, the
successive application of the renormalization procedure drives the
system away from the fixed point. On the other hand at $T=T^*$ the
system remains at the fixed point under the application of the RG
procedure. \label{fig3} }
\end{figure}

To proceed in the determination of these properties let us have a
closer look at the partition function and at the free energy of
the system. First, after taking a partial trace of the partition
function $Z_N\{{\cal H_{\rm eff}}\}$ we get the Boltzmann
distribution in $N^{\prime}$ spin variables $S^{\prime}$:
\begin{equation}
\re^{-{\cal H}_{\rm eff}^{\prime}(S^{\prime})} = Sp_{N^{\prime
\prime}}^{S^{\prime \prime}} \re^{-{\cal H}_{\rm eff}(S)},
\label{12.40}
\end{equation}
where $Sp_{N^{\prime \prime}}^{S^{\prime \prime}}$ stands for the
trace over $N^{\prime \prime}= N - N^{\prime}$ spin variables
$S^{\prime \prime}$. Now taking trace over the remaining $N^{\prime}$
spin variables one gets:
\begin{equation}
Z_{N^{\prime}}\{{\cal H}_{\rm eff}^{\prime}\} =
Sp_{N^{\prime}}^{S^{\prime}} \re^{-{\cal H}_{\rm
eff}^{\prime}(S^{\prime})}. \label{12.41}
\end{equation}
Substituting into the right hand side of (\ref{12.41}) its explicit
form given by (\ref{12.40}) one gets:
\begin{equation}
\nonumber Z_{N^{\prime}}\{{\cal H}_{\rm eff}^{\prime}\} =
Sp_{N^{\prime}}^{S^{\prime}} Sp_{N^{\prime \prime}}^{S^{\prime
\prime}}  \re^{-{\cal H}_{\rm eff}(S)}= Z_{N}\{{\cal H}_{\rm eff}
\}.
\end{equation}
This relation can be rewritten in terms of the  {\it flow
equations} which describe a motion of a point describing an effective
Hamiltonian:
\begin{eqnarray}
K^{\prime} &=& {\cal R}_K (K,h),
\label{12.43} \\
h^{\prime} &=& {\cal R}_h (K,h),
\label{12.44} \\
C^{\prime} &=& b^d C + {\cal R}_0 (K,h),
\label{12.45}
\end{eqnarray}
for the {\em couplings} defining ${\cal H}_{\rm eff}$. As far as
the temperature $T$ enters parameter $K$, the above relations
imply the flow equations for $T$ as well. To study this let us
consider $H=0$ and write the flow equation for the temperature as:
\begin{equation}
T \Rightarrow T^{\prime}= {\cal R}(T),
\label{12.46}
\end{equation}
with ${\cal R}(T)$ being appropriate function of $T$. Suppose that
${\cal R}(T)$ has a form given in the figure~\ref{fig3}. The
important feature there is that ${\cal R}(T)$ crosses the line
$T^{\prime}=T$ at some point $T=T^*$. This point is called the
{\it fixed point} (FP). It is clear that when $T$ is smaller or
greater than $T^*$ the successive application of the
renormalization procedure drives the system away from the fixed
point. On the other hand at $T=T^*$ the system remains at the
fixed point under the application of the RG procedure. Recalling
the flow equation for the correlation length (\ref{12.39}) which
can be written as:
\begin{equation}
\xi(T) = b \xi(T^{\prime}),
\label{12.47}
\end{equation}
one has that in the FP this reads:
\begin{equation}
\xi(T^*) = b \xi(T^*).
\label{12.48}
\end{equation}
Since $b>1$ it is possible only when
\begin{itemize}
\item $\xi(T^*) = \infty$ and
\item $\xi(T^*) = 0$.
\end{itemize}
The first case characterizes the critical point
$T^*=T_{\mathrm{c}}$, whereas the second corresponds to the
vanishing of the correlation length at zero or infinite
temperatures (when the spins are frozen at a ground state or
totally uncoupled). Note, that this corresponds to two FPs more
($T^*=0$, $T^*=\infty$) in the figure~\ref{fig3}.

Having identified the fixed point $T^*$ of the RG transformation
as the critical point of the system let us study what knowledge
about the critical exponents can be obtained based on the
properties of the RG transformation. First let us linearize it in
the vicinity of the FP introducing the variable:
\begin{equation}
\tau=\frac{T-T_{\mathrm{c}}}{T_{\mathrm{c}}} = \frac{T-T^*}{T^*}
 \label{12.49}
\end{equation}
and replacing the plot of ${\cal R}(T)$ near $T^*$ by its tangent
at $T^*$.

Then after the renormalization the temperature deviation will be:
\begin{equation}
\tau^{\prime} \equiv \tau^{(1)} \simeq \Lambda_1(b) \tau
\label{12.50}
\end{equation}
for small enough $\tau$, where $\Lambda_1(b)$  is the slope of the
tangent. To find its dependence on $b$ let us apply the
renormalization procedure twice:
\begin{equation}
\tau^{\prime \prime} \equiv \tau^{(2)}  \simeq \Lambda_1(b)
\Lambda_1(b) \tau .
\label{12.51}
\end{equation}
Such transformation should be equivalent to
transforming with a spatial rescaling factor $b^2$:
\begin{equation}
\tau^{(2)}  \simeq \Lambda_1(b^2)  \tau ,
\label{12.52}
\end{equation}
or
\begin{equation}
\Lambda_1(b) \Lambda_1(b) = \Lambda_1(b^2).
\label{12.53}
\end{equation}
This leads to the conclusion that $\Lambda_1(b)$ should have the
following form:
\begin{equation}
\Lambda_1(b) = b^{\lambda_1},
\label{12.54}
\end{equation}
with $\lambda_1$ being constant independent of $b$. Now for the
correlation length after $l$ transformations one gets:
\begin{equation}
\xi(\tau) = b^l\xi([\Lambda_1(b)]^l \tau) =
b^l\xi(b^{l\lambda_1}\tau).
\label{12.55}
\end{equation}
As far as (\ref{12.55}) holds for any $l$  (and for small $\tau$(!))
let us choose it to satisfy:
$$
b^{l\lambda_1}\tau =1,
$$
or
\begin{equation}
b^l = (1/\tau)^{1/\lambda_1}.
\label{12.56}
\end{equation}
Then from (\ref{12.55}) one gets:
\begin{equation}
\nonumber
\xi(\tau) = (1/\tau)^{1/\lambda_1} \xi(1)
\end{equation}
or
\begin{equation}
\xi(\tau) \sim \tau^{-1/\lambda_1}. \label{12.57}
\end{equation}
Comparing (\ref{12.57}) with the definition of the critical exponent
$\nu$: $\xi(\tau) \sim \tau^{-\nu}$ we get:
\begin{equation}
\nu =1/\lambda_1.
\label{12.58}
\end{equation}
So as we have seen, the knowledge of the linearized RG transformation
enables one to determine the critical exponent!

Note, that having carried out similar analysis for the free energy
per spin, $f={-}{N^{-1}}\ln Z_N$, one arrives at the expression: $
f(\tau) =b^{-d l}f(b^{\lambda_1 l}\tau)$. Again, choosing $b$
from (\ref{12.56}) one gets $f(\tau)
=\tau^{d/\lambda_1}f(1)=\tau^{d \nu}f(1)$. Comparing the
latter with the scaling behaviour of the free energy in the
vicinity of the critical point $f(\tau)\sim \tau^{2-\alpha}$ one
proves the {\em hyperscaling relation}:
\begin{equation}
2-\alpha= d\nu.
 \label{12.61}
\end{equation}

Similarly, considering the case of non-zero magnetic field one can
write the recursion relations for the renormalized temperature and
field. Their linearization in the vicinity of the fixed point
leads to the {\em matrix of the linear RG operator} with two
eigenvalues, defining two different critical exponents.

To complete the RG study of the 1d Ising model it remains to
perform the above described procedure explicitly and to find the
values of the critical exponents (at zero-temperature FP $T=H=0$,
\cite{Fisher}). The above described transformations concisely give
the main features of the RG transformation, allowing us, in
particular, to define the critical exponents of the system.

\section{From the spin Hamiltonians to the effective ones} \label{V}

Very often a starting point for the RG study of critical behaviour
of a many-particle system is its {\em effective Hamiltonian}.
Taken, that in the  model description the system  is identified
with the {\em spin Hamiltonian} (see section~\ref{III}), the
effective Hamiltonian arises as a certain ``metamodel''. It shares
global features of different spin Hamiltonians: their
dimensionalities, symmetries, type of interparticle interaction
and as a result, brings about common features of their critical
behaviour. Below, we shall show how to obtain the effective
Hamiltonians for the spin models of section~\ref{III}.

\subsection{$m$-vector model}\label{Va}
To get an effective Hamiltonian we proceed as follows. Let us
define the free energy ${\cal F}$ and the partition function
${\cal Z}$ of a spin model (\ref{3.1})  as:
\begin{equation} \label{5.2}
{\cal F} = -\beta^{-1} \ln {\cal Z}, \qquad {\cal Z}={\rm Sp}
\re^{-\beta {\cal H}},
\end{equation}
where $\beta=(k_{\mathrm{B}}T)^{-1}$ and ${\rm Sp}(\dots) $ as
usually means the sum over all possible states. In our case it
corresponds to the integration over the surface of $m$-dimensional
hypersphere (we take it to be of unit radius):
\begin{equation} \label{5.3}
{\rm Sp} (\dots) = \prod_{\bf R} \int {\rm d} \vec{S}_{\bf R}
\delta (|\vec{S}_{\bf R}| - 1) (\dots).
\end{equation}
Let us introduce the Fourier-transforms of the variables
$\vec{S}_{{\bf R}}$ by:
\begin{equation} \label{5.4}
\vec{S}_{{\bf R}} = \frac{1}{\sqrt N} \sum_{{\bf k}} \re^{\ri{\bf
k}{\bf R}} \vec{S}_{{\bf k}}\,, \hspace{1cm} \vec{S}_{{\bf k}} =
\frac{1}{\sqrt N} \sum_{{\bf R}} \re^{-\ri{\bf k}{\bf R}}
\vec{S}_{{\bf R}}\,.
\end{equation}
Here and below, when it will not be mentioned explicitly, vector
${\bf R}$  spans all sites of the lattice whereas ${\bf k}$
changes in the first Brillouin zone. Now the Hamiltonian
(\ref{3.1}) can be rewritten as :
\begin{equation} \label{5.5}
{\cal H} =  -\frac{1}{2} \sum_{{\bf k}} \nu(k) \vec{S}_{{\bf k}}
\vec{S}_{-{\bf k}}\,,
\end{equation}
where we have introduced Fourier transform of the potential of
interaction by:
\begin{equation} \label{5.6}
J(R) =  \frac{1}{ N} \sum_{{\bf k}} \re^{\ri{\bf k}{\bf R}}
\nu(k), \hspace{1cm} \nu(k) =  \sum_{{\bf R}} \re^{-\ri{\bf k}{\bf
R}} J(R).
\end{equation}
Now the partition function (\ref{5.2}) reads:
\begin{equation} \label{5.7}
{\cal Z} ={\rm Sp} \re^{\frac{\beta }{2} \sum_{{\bf k}} \nu(k)
\vec{S}_{{\bf k}} \vec{S}_{-{\bf k}}}.
\end{equation}
In order to take the trace in (\ref{5.7}) let us transform it to
the expression with linear dependence on $\vec{S}_{{\bf k}}$. To
this end, one makes use of the Stratonovich-Hubbard transformation
introducing the field variable $\vec{\phi}_{{\bf k}}$ which is
conjugated to the spin variable $\vec{S}_{{\bf k}}$ by the
identity:
\begin{eqnarray} \nonumber
 \re^{-\beta {\cal H}} &=& \prod_{\vec{k}} \re^{\frac{\beta
\nu(k)}{2}\vec{S}_{{\bf k}} \vec{S}_{-{\bf k}}} = \prod_{\vec{k}}
(\frac{1}{2\pi \beta \nu(k)})^{m/2} \int {\rm d} \vec{\phi}_{{\bf
k}} \re^{\frac{-1}{2\beta \nu(k)} \vec{\phi}_{{\bf k}}
\vec{\phi}_{{\bf -k}} + \vec{S}_{{\bf k}}  \vec{\phi}_{{\bf -k}}}
\\
&&{}\sim{}  \int ({\rm d} \vec{\phi}) \re^{\sum_{\bf k}(
\frac{-1}{2\beta \nu(k)} |\vec{\phi}_{{\bf k}}|^2 + \vec{S}_{{\bf
k}} \vec{\phi}_{{\bf -k}})}.  \label{5.8}
\end{eqnarray}
As far as the Stratonovich-Hubbard transformation was performed
for each function $\vec{S}_{{\bf k}}$ we arrived at the functional
integral $\int ({\rm d} \vec{\phi}) = \prod_{{\bf R}}
\prod_{i=1}^m \int_{-\infty}^{\infty} {\rm d} \phi^{(i)}_{{\bf
R}}$ over the field variables
\begin{equation} \label{5.10}
\vec{\phi}_{{\bf R}} = \frac{1}{\sqrt N} \sum_{{\bf k}}
\re^{\ri{\bf k}{\bf R}} \vec{\phi}_{{\bf k}}\,, \hspace{1cm}
\vec{\phi}_{{\bf k}} = \frac{1}{\sqrt N} \sum_{{\bf R}}
\re^{-\ri{\bf k}{\bf R}} \vec{\phi}_{{\bf R}}\,.
\end{equation}
From now on we omit the coefficients in front of the functional
integral.

We have reached our goal: now the trace in the expression for the
partition function concerns only the last term of the exponents
under integration in (\ref{5.7}) which is a linear function of
$\vec{S}$:
\begin{equation} \label{5.11}
{\cal Z} = {\rm Sp} \re^{\frac{\beta}{2} \sum_{{\bf k}} \nu(k)
\vec{S}_{{\bf k}} \vec{S}_{-{\bf k}}} \sim \int ({\rm d}
\vec{\phi}) \re^{\frac{-1}{2} \sum_{{\bf k}} \frac{1}{\beta \nu
(k)} |\vec{\phi}_{{\bf k}}|^2} {\rm Sp} \,  \re^{\sum_{{\bf
R}}\vec{S}_{{\bf R}} \vec{\phi}_{{\bf R}}}.
\end{equation}
The last step is to take integral over $\vec{S}_{{\bf R}}$ (recall
that the trace is defined by equation~(\ref{5.3})). This can be
achieved by passing from the $m$-dimensional Cartesian coordinates
$S^{1}, S^{2}, \dots, S^{m}$ to the $m$-dimensional polar ones $S,
\theta_1, \dots ,\theta_{m-1}$. Performing this integration and
representing the result as a series in $\vec{\phi}$ one gets for
the partition function:
\begin{equation} \label{5.12}
{\cal Z} \sim \int ({\rm d} \vec{\phi})\re^{\frac{-1}{2}
\sum_{{\bf k}} (\frac{1}{\beta \nu(k)} - u_2) |\vec{\phi}_{{\bf
k}}|^2 - \sum_{{\bf R}} \sum_{l=2}^{\infty}
\frac{u_{2l}}{(2l)!}|\vec{\phi}_{{\bf R}}|^{2l}},
\end{equation}
where the coefficients $u_{2l}$ readily follow: $u_2 = -1/m$,
$u_4=6/m^2(m+2)$, $\dots$ Expression (\ref{5.12}) gives the
functional representation of the partition function of the
$m$-vector model. So far we have not gained a lot by the above
described transformations: the difficulty of taking ${\rm Sp}$ of
the initial expression (\ref{5.5}) now is transformed into the
difficulty of calculating a functional integral (\ref{5.12}).
However, in order to study critical behaviour of the model
(\ref{3.1}), expression (\ref{5.12}) can be further approximated.
It appears that all of $\phi$ higher than the fourth powers  do
not effect the asymptotic critical behaviour at $d=3$: they do not
change the value of the fixed point of RG transformation (cf.
section \ref{IV}) and are {\em irrelevant} in the RG sense
\cite{Wilson,RGbooks}. Therefore one can be restricted to the
$\phi^4$ model. Being interested in the long-range correlations
arising in the system in the vicinity of a critical point, one
substitutes the Fourier image of the interaction potential by its
expansion for small wave vector values $k$:
$\nu(k)\simeq\nu(0)-{1}/{2}|\nu''(0)|k^2$ and writes the
contribution $\sum_{\bf k} k^2 \vec \phi_{\bf k} \vec \phi_{-\bf
k}$ in the form $\sum_{\bf R} (\nabla \vec \phi_{\bf R})^2$
re-scaling variables $\phi$ to get the gradient term without any
coefficient. One further passes to the continuous limit $\sum_{\bf
R} \rightarrow \int {\rm d} {\bf R}$ and gets for the free energy:
\begin{equation}\label{5.13}
{\cal F} \sim \ln \int({\rm d}\vec{\phi}) \re^{-{\cal H}_{\rm
eff}}
\end{equation}
with the {\em effective Hamiltonian}:
\begin{equation} \label{5.14}
{\cal H}_{\rm eff} = \int {\rm d}^d R \left\{ \frac{1}{2}
\left((\nabla \phi)^2 +\mu_0^2 \phi^2\right) + \frac{u_0}{4!}
\phi^4\right\},
\end{equation}
where $\phi\equiv\vec{\phi}_{\bf R}$,
$\phi^2\equiv|\vec{\phi}_{\bf R}|^2$, variables $\mu_0$ and $u_0$
are called bare mass and coupling. From the above derivation we
know that $u_0$ is positive ($u_0\sim u_4$ from
equation~(\ref{5.12})), which ensures the existence of the
integral (\ref{5.13}). Effective Hamiltonian (\ref{5.14}) shares
common global properties with the spin Hamiltonian (\ref{3.1}):
the dimension of space (i.e. dimension of vectors ${\bf k}$, ${\bf
R}$), dimension of the order parameter ($m$)  and its symmetry:
the functional representation we obtained is symmetric under group
of rotations in the $m$-dimensional space $O(m)$: it depends only
on $|\vec{\phi}|$. Let us now see what differences will appear in
the functional representation of the non-ideal models.

\subsection{Cubic model}\label{Vb}

It is intuitively clear that an effective Hamiltonian of the cubic
model (\ref{3.2}) will differ from that of the $m$-vector model
(\ref{5.14}) by the presence of terms with new, cubic symmetry.
Indeed this is the case that one can easily check following the
next steps: starting from the Hamiltonian (\ref{3.2}) via the
Stratonovich-Hubbard transformation one obtains the functional
representation for the interaction part (which coincides with
equation~(\ref{5.8})). An additional term $\re^{-\beta V \sum_i
S^i}$ can be represented in a form of the functional derivative
$\re^{-\beta V \sum_i \partial^4 / \partial (\phi^i)^4}$,
resulting in the following expression for the partition function
of the cubic model (c.f. equation~(\ref{5.11})):
\begin{equation} \label{5.14a}
{\cal Z}  \sim \int ({\rm d} \vec{\phi}) \re^{\frac{-1}{2}
\sum_{{\bf k}} \frac{1}{\beta \nu (k)} |\vec{\phi}_{{\bf k}}|^2}
\re^{-\beta V \sum_{\bf R} \sum_i \partial^4 / \partial
(\phi^i_{{\bf R}})^4} {\rm Sp} \,  \re^{\sum_{{\bf
R}}\vec{S}_{{\bf R}} \vec{\phi}_{{\bf R}}}.
\end{equation}
Taking trace in equation~(\ref{5.14a}) leads  to the familiar
expression (\ref{5.12}). However, the  derivative $\partial^4 /
\partial (\phi^i_{\bf R})^4$ gives rise to the contributions of
cubic symmetry, the lowest order contribution being proportional
to $v_0 \sum_i (\phi^i_{\bf R})^4$. The resulting effective
Hamiltonian reads:
\begin{equation} \label{5.14b}
{\cal H}_{\rm eff} = \int {\rm d}^d R \left\{ \frac{1}{2}
\left((\nabla \phi)^2 + \mu_0^2 \phi^2\right) + \frac{u_0}{4!}
\phi^4 + \frac{v_0}{4!} \sum_{i=1}^m (\phi^i)^4 \right\},
\end{equation}
and contains two bare couplings $u_0$ and $v_0$, corresponding to
the $\phi^4$ terms of different symmetries. Coupling $u_0$ is
positive, whereas the sign of the coupling $v_0$ coincides with
that of $V$ in equation~(\ref{3.2}). At $v_0=0$ one recovers an
effective Hamiltonian of the $m$-vector model (\ref{5.14}).

\subsection{Weakly diluted quenched $m$-vector model}\label{Vc}

The peculiarities of the free energy calculation for the model
(\ref{3.3}) consist in averaging over quenched disorder. Indeed,
for each configuration of empty and occupied sites in (\ref{3.3})
one can write a corresponding configuration-dependent partition
function $Z_{{\rm conf}}$:
\begin{equation} \label{5.15}
{\cal Z}_{{\rm conf}} = {\rm Sp}_{\vec{S}} \re^{-\beta \cal{H}},
\end{equation}
where ${\rm Sp}_{\vec{S}}$ concerns spin degrees of freedom and is
defined by (\ref{5.3}). The free energy is obtained as the
configurational average:
\begin{equation} \label{5.16}
{\cal F}= -\beta^{-1} \langle \ln {\cal Z}_{{\rm
conf}}\rangle_{{\rm conf}}.
\end{equation}
One of the ways to proceed is to make use of the replica trick
\cite{replicas}, which allows us to avoid integration of the
logarithm in (\ref{5.16}) substituting it by a power function:
\begin{equation} \label{5.17}
\ln {\cal Z} = \lim_{n \rightarrow 0} \frac{{\cal Z}^n -1}{n}\,.
\end{equation}
Then $Z_{{\rm conf}}^n$ can be written as:
\begin{equation} \label{5.18}
{\cal Z}_{{\rm conf}}^n = {\rm Sp} \re^{\frac{\beta}{2}\sum_{\bf
R,R'} J(|{\bf R} - {\bf R'}|) \sum_{\alpha=1}^n
\vec{\sigma}^{\alpha}_{{\bf R}} \vec{\sigma}^{\alpha}_{{\bf R'}}}.
\end{equation}
with obvious notations ${\rm Sp} (\dots) = \prod_{\alpha=1}^n {\rm
Sp}_{\vec{S}^{\alpha}} (\dots)$ and $\vec{\sigma}^{\alpha}_{{\bf
R}} \equiv c_{{\bf R}} \vec{S}^{\alpha}_{{\bf R}}$. Upcoming
calculations closely follow lines of the subsection \ref{Va}.
Introducing by the Stratonovich-Hubbard transformation field
variables $\vec{\phi}^{\alpha}_{{\bf R}}$, conjugated to
$\vec{\sigma}^{\alpha}_{{\bf R}}$ one is able to take the trace
over spin subsystem and is left with the configuration-dependent
partition function:
\begin{equation} \label{5.19}
{\cal Z}_{\rm conf}^n \sim \int ({\rm d} \vec{\phi})
\re^{\frac{-1}{2} \sum_{{\bf k}}  \frac{1}{\beta \nu
(k)}\sum_{\alpha=1}^n \vec{\phi}_{{\bf k}}^{\alpha}
\vec{\phi}_{-{\bf k}}^{\alpha} - \sum_{{\bf R}}
\sum_{l=1}^{\infty} \frac{u_{2l}}{(2l)!} \sum_{\alpha=1}^n
|\vec{\phi}^{\alpha}_{{\bf R}}|^{2l} c_{{\bf R}}}.
\end{equation}
Note that $c_{{\bf R}}$ appears in  (\ref{5.19}) in the first
power because from $c_{{\bf R}}=\{0,1\}$ it follows that $(c_{{\bf
R}})^{l} \equiv c_{{\bf R}}$. The last step is to perform
configurational averaging of (\ref{5.19}) with the distribution
function (\ref{3.5}). Let us represent the result of averaging in
the exponential form
\begin{equation}\label{5.20}
\left\langle \re^{-\sum_{{\bf R}} \sum_{l=1}^{\infty}
\frac{u_{2l}}{(2l)!} \sum_{\alpha=1}^n |\vec{\phi}^{\alpha}_{{\bf
R}}|^{2l} c_{{\bf R}}} \right\rangle_{\rm conf} = \prod_{{\bf R}}
\re^{\sum_{p \geqslant  1} \frac{(-1)^p}{p!} \kappa_p(c) \Big (
\sum_{l=1}^{\infty} \frac{u_{2l}}{(2l)!} \sum_{\alpha=1}^n
|\vec{\phi}^{\alpha}_{{\bf R}}|^{2l} \Big )^p}\,,
\end{equation}
where $\kappa_p(c)$ are cumulants of random variables $c_{{\bf
R}}$ and can be easily calculated for the random variable
distribution (\ref{3.5}):
\begin{eqnarray}
\nonumber \kappa_1 &=& \sum_{c_{{\bf R}}=\{0,1\}} c_{{\bf R}}
p(c_{{\bf R}})=c,
\\ \label{5.21}
\kappa_2 &=& \sum_{c_{{\bf R}}=\{0,1\}} (c_{{\bf R}})^2 p(c_{{\bf
R}}) - \left(\sum_{c_{{\bf R}}=\{0,1\}} c_{{\bf R}} p(c_{{\bf
R}})\right)^2 =c(1-c), \dots .
\end{eqnarray}
For the free energy one gets:
\begin{equation}
{\cal F}= -\beta^{-1} \langle \ln Z_{{\rm conf}}\rangle_{{\rm
conf}}= -\beta^{-1} \lim_{n \rightarrow 0} \left\{ \frac{c_1}{n}
\int({\rm d} \vec{\phi})^n \re^{-F[\phi]} -1/n \right\},
\label{5.22}
\end{equation}
where the ($n$-dependent) coefficient $c_1$ can be recast
explicitly following all the steps of calculations described above
and the free energy functional $F[ \phi ]$ is given by:
\begin{equation}
F[ \phi ] =  \frac{1}{2}\sum_{{\bf k}} \sum_{\alpha=1}^n
\frac{1}{\beta \nu(k)} |\vec{\phi}^{\alpha}_{{\bf k}}|^2 - \sum_{p
\geqslant  1} \frac{(-1)^p}{p!} \kappa_p(c) \sum_{{\bf R}} \left(
\sum_{l=1}^{\infty} \frac{u_{2l}}{(2l)!} \sum_{\alpha=1}^n
|\vec{\phi}^{\alpha}_{{\bf R}}|^{2l} \right)^p. \label{5.23}
\end{equation}
Again, as in the previous subsection, restricting ourselves to the
$\phi^4$ terms, expanding the short-range interaction potential,
re-scaling the fields and passing to the continuous limit one gets
for the free energy:
\begin{equation} \label{5.24}
{\cal F} \sim \int({\rm d}\vec{\phi}) \re^{-{\cal H}_{\rm eff}},
\end{equation}
now the proportionality sign also hides, besides the coefficient,
 the replica limit, as written explicitly in (\ref{5.22}). The
effective Hamiltonian reads:
\begin{equation} \label{5.25}
{\cal H}_{\rm eff} = \int {\rm d}^d R \left\{ \frac{1}{2}
\sum_{\alpha=1}^n \left[(\nabla \phi^\alpha)^2+\mu_0^2
(\phi^\alpha)^2\right] + \frac{u_0}{4!} \sum_{\alpha=1}^n
(\phi^\alpha)^4 + \frac{v_0}{4!} \sum_{\alpha,\beta=1}^n
(\phi^\alpha)^2 (\phi^\beta)^2 \right\}.
\end{equation}
Here, the coupling $u_0$ is positive (being proportional to
$cu_4$, (\ref{5.12})) whereas the coupling $v_0$ is proportional
to $c(c-1)u_2^2$ and is negative. The last term in (\ref{5.25}) is
present only for non-zero dilution: it is directly responsible for
the effective interaction between replicas due to the presence of
impurities.

\subsection{Random anisotropy model}\label{Vd}

To treat the random anisotropy in the spin Hamiltonian
(\ref{3.6}), one first writes the configuration-dependent
partition function for a fixed local anisotropy axes configuration
$\{\hat x\}$:
\begin{equation}\label{5.26}
{\cal Z}_{\rm conf}(\{\hat x\})={\rm Sp}_{\vec{S}}\re^{-\beta
{\cal H}}.
\end{equation}
As in equation~(\ref{5.15}), trace in equation~(\ref{5.26})
concerns only the spin degrees of freedom. Applying the
Stratonovich-Hubbard transformation to the interaction part of the
Hamiltonian one presents (\ref{5.26}) in the form of the
functional integral and is able to  take the trace:
\begin{equation}\label{5.27}
{\cal Z}_{\rm conf}\left(\{\hat x\}\right)\sim\int(\rd\vec{\phi})
\re^{-{\cal H}(\hat x)},
\end{equation}
with
\begin{equation} \label{5.28}
{\cal H}({\hat x}) = \int{\rm d}^d R \Big \{
\frac{1}{2}\left[(\nabla \phi)^2 + \mu_1^2 \phi^2\right]-D_1 (\phi
\hat x)^2+v_1\phi^4+ z_1 \phi^2(\phi \hat x)^2+ \cdots \Big \} .
\end{equation}
Again, the expansion for small $k$ was performed and the
continuous limit has been taken. Explicit values for the
coefficients $\mu_1$, $D_1$, $v_1$, $z_1$ are given in
\cite{Dudka05}.

The rest of calculations follow the steps outlined in the section
\ref{Vc}: for the quenched disorder the free energy is defined by
(\ref{5.16}), where averaging is performed over the random axis
distribution (given by (\ref{3.7}) or (\ref{3.8})). Subsequently,
one substitutes $\ln {\cal Z}$ by a power function via the replica
trick (\ref{5.17}). For the isotropic random axis distribution
 (\ref{3.7}) one gets the effective Hamiltonian \cite{Aharony75}
containing three $\phi^4$ couplings of different symmetry:
\begin{equation}\label{5.29}
{\cal H}_{\rm eff} = \int {\rm d}^d R \left\{ \frac{1}{2}\left[
(\nabla \varphi)^2 + \mu_0^2 \varphi^2 \right]+u_0\varphi^4+
v_0\sum_{\alpha=1}^n (\phi^{\alpha})^4+
w_0\sum_{\alpha,\beta=1}^n\sum_{i,j=1}^m
\phi_i^{\alpha}\phi_j^{\alpha}\phi_i^{\beta}\phi_j^{\beta} \right
\}.
\end{equation}
Here and below $\varphi^2=\sum_{\alpha=1}^n(\phi^\alpha)^2$.  One
can check the signs of the couplings:
 $u_0\sim \frac{D^2}{m^2(m+2)}>0$, $v_0\sim D>0$,
$w_0\sim \frac{-D^2}{m(m+2)}<0$. Moreover, from the explicit
expressions for the couplings $w_0$ and $u_0$ one gets for their
ratio $w_0/u_0=-m$. The latter relation determines a region of
physically allowed initial values in the $(u-v-w)$-space of
couplings.

For the cubic distribution (\ref{3.8}) the average over the random
variables $\{\hat x\}$ leads to the effective Hamiltonian
\cite{Aharony75} with four couplings:
\begin{eqnarray} \nonumber
 {\cal H}_{\rm eff}&=& \int{\rm d}^d R \left\{
 \vphantom{ w_0\sum_{i=1}^m\sum_{\alpha,\beta{=}1}^n
\left(\phi_i^{\alpha}\right)^2\left(\phi_i^{\beta}\right)^2}
  \frac{1}{2}\left[(\nabla\varphi)^2 + \mu_0^2\varphi^2
\right]+u_0\varphi^4+v_0\sum_{\alpha=1}^n(\phi^{\alpha})^4\right.
\\ &&{}\left. \label{5.30}+
 w_0\sum_{i=1}^m\sum_{\alpha,\beta{=}1}^n
\left(\phi_i^{\alpha}\right)^2\left(\phi_i^{\beta}\right)^2+
y_0\sum_{i=1}^m\sum_{\alpha=1}^n\left(\phi_i^{\alpha}\right)^4
\right\}.
\end{eqnarray}
The signs of the couplings are as follows: $u_0 \sim
{D^2}/{(2m^2)}>0$, $v_0\sim D>0$, $w_0\sim {-D^2}/{(2m)}<0$. The
last term in (\ref{5.30})  is of cubic symmetry. It has to be
included since it is generated if one further applies the RG
transformation. Therefore $y_0$ can be of either sign. The
symmetries of $w_0$ terms in (\ref{5.29}) and (\ref{5.30}) differ.
However the ratio $w_0/u_0=-m$ holds for the effective Hamiltonian
(\ref{5.30}) as well.

\subsection{Stacked triangular antiferromagnet}\label{Ve}

It is an interaction potential (\ref{3.9}) that makes a difference
in a derivation of the effective Hamiltonian for the frustrated
magnet and the regular one. One can find the details of the
procedure to derive an effective Hamiltonian via the
Stratonovich-Hubbard transformation in \cite{Kawamura88}. The main
difference from the above described calculations for the
$m$-vector magnet (subsection \ref{Va}) arises when one proceeds
with the Gaussian term in equation~(\ref{5.12}):
\begin{equation} \label{5.31}
\sum_{{\bf k}} \left(\frac{1}{\beta \nu(k)} - u_2\right)
|\vec{\phi}_{{\bf k}}|^2 \equiv \sum_{{\bf k}} d(k)
|\vec{\phi}_{{\bf k}}|^2.
\end{equation}
The sum over ${\bf k}$ in (\ref{5.31}) spans the 1st Brillouin
zone. For the $m$-vector model, $d(k)$ has a minimum at $k=0$ and
an expansion in $k$ around this minimum results in the effective
Hamiltonian (\ref{5.14}). Note, that one minimum in $d(k)$ leads
to one field variable $\phi_{\bf k}$ in the effective Hamiltonian.
For the frustrated model, the $d(k)$ has a maximum at $k=0$ and
two independent minima at $k\neq 0$ in the 1st Brillouin zone.
Subsequently, the zone can be rearranged into separate subzones
and the field $\phi_{\bf k}$ over the original Brillouin zone can
be decomposed into two fields, $\phi_{1,{\bf k}}$ and
$\phi_{2,{\bf k}}$, with ${\bf k}$ confined to the new subzone.
Now, the minima of $d(k)$ occur at the origin and pertain to the
fields $\phi_{1}$ and $\phi_{2}$. The above decomposition results
in the appearance of the terms of new symmetry in the effective
Hamiltonian, which now reads:
\begin{equation} \label{5.32}
{\cal H}_{\rm eff} \!=\! \int{\rm d}^d R \left\{
\frac{1}{2}\left[\mu_0^2 (\phi_1^2+\phi_2^2) +(\nabla\phi_1)^2+
(\nabla\phi_2)^2 \right] +
\frac{u_0}{4!}\left[\phi_1^2+\phi_2^2\right]^2 +
\frac{v_0}{4!}\left[(\phi_1 \cdot \phi_2)^2- \phi_1^2\phi_2^2
\right] \right \},
\end{equation}
with a scalar product of $m$-component fields $\phi_1 \cdot \phi_2
= \sum_{i=1}^m\phi_1^i \phi_2^i$. The coupling $u_0$ in
(\ref{5.32}) is positive, whereas the sign of the coupling $v_0$
determines a type of spin ordering: a non-collinear order occurs
for $v_0>0$.  For $v_0<0$ the fields $\phi_1$ and $\phi_2$ tend to
allign parallel or antiparallel, this corresponds to the
sinusoidal ordering or the linearly-polarized spin-density wave.

\section{RG explanation of
criticality in non-ideal systems} \label{VI}

Now with the effective Hamiltonians (section \ref{V}) and the RG
method (section \ref{IV}) at hand one can proceed further,
applying the method in order to study the critical behaviour of
the models. However, to be honest with the reader, he should be
warned that the whole story is not that simple as shown in the
section \ref{IV}. It is not only because the very models we are
interested in are much more complicated as compared to the 1d
Ising model: in what follows below we will be interested in the
criticality of 3d systems. The {\em real space renormalization}
described in the section \ref{IV} works for the low-dimensional
systems, whereas for realistic 3d systems one rather applies the
RG equations in the momentum space. Moreover, the very approach to
RG as to the way of calculating the partition function of the
system \cite{Wilson} has its alternative known as the
field-theoretical RG \cite{RGbooks}. The latter is a tool to cope
with the divergencies of correlation functions (vertex functions).
Here, the RG procedure consists in the controlled rearrangement of
the perturbation theory expansions giving rise to the RG
equations. However, the underlying notions of both procedures are
the same: given the effective Hamiltonian one applies the RG
transformation and studies the flow equations for the couplings of
the Hamiltonian. If the flow equations possess the fixed point
(FP), which is stable and reachable from the initial conditions,
it corresponds to the critical point of the system. Scaling arises
in the vicinity of this FP and the universal values of the
exponents governing scaling of different physical quantities may
be found. Having this preamble in  mind, we can make a brief
review of what results for the RG flows, FPs and exponents of the
non-ideal models have been found so far and how  they differ from
those of the ``ideal'' $m$-vector model.

\subsection{$m$-vector magnet}\label{VIa}

Put in a formal way, the question: does the 3d $m$-vector magnet
exhibit a critical point (a Curie point for a ferromagnet or a
N\'eel point for an antiferromagnet), transforms into the
question: is there a stable and reachable FP in the flow equations
(cf. (\ref{12.43})--(\ref{12.45})) for the couplings of its
effective Hamiltonian? The flow equation for the coupling $u$ of
the effective Hamiltonian (\ref{5.14}) can be written in a general
form of an ordinary first order differential equation:
\begin{equation} \label{6.1}
\frac{\rd u}{\rd\ln \ell}=\beta(u),
\end{equation}
with the RG {\em flow parameter} $\ell$ and {\em $\beta$-function}
$\beta(u)$. The parameter $\ell$ may serve to measure a distance
to the critical point: $\ell\rightarrow 0$ corresponds to
$T\rightarrow T_{\mathrm{c}}$, and a specific form of the
$\beta$-function depends on a choice of the RG procedure. The data
discussed in this chapter have been obtained within the
field-theoretic RG \cite{RGbooks} approach at $d=3$. Giving
numbers, we shall refer to the source, where they were obtained.
However we shall try to be not too specific, mentioning several
review papers for the interested reader.

\begin{figure}[h]
\begin{picture}(80,170)
\put(20,20){\parbox[t]{110mm}{\includegraphics[width=
60mm,height=50mm,angle=0]{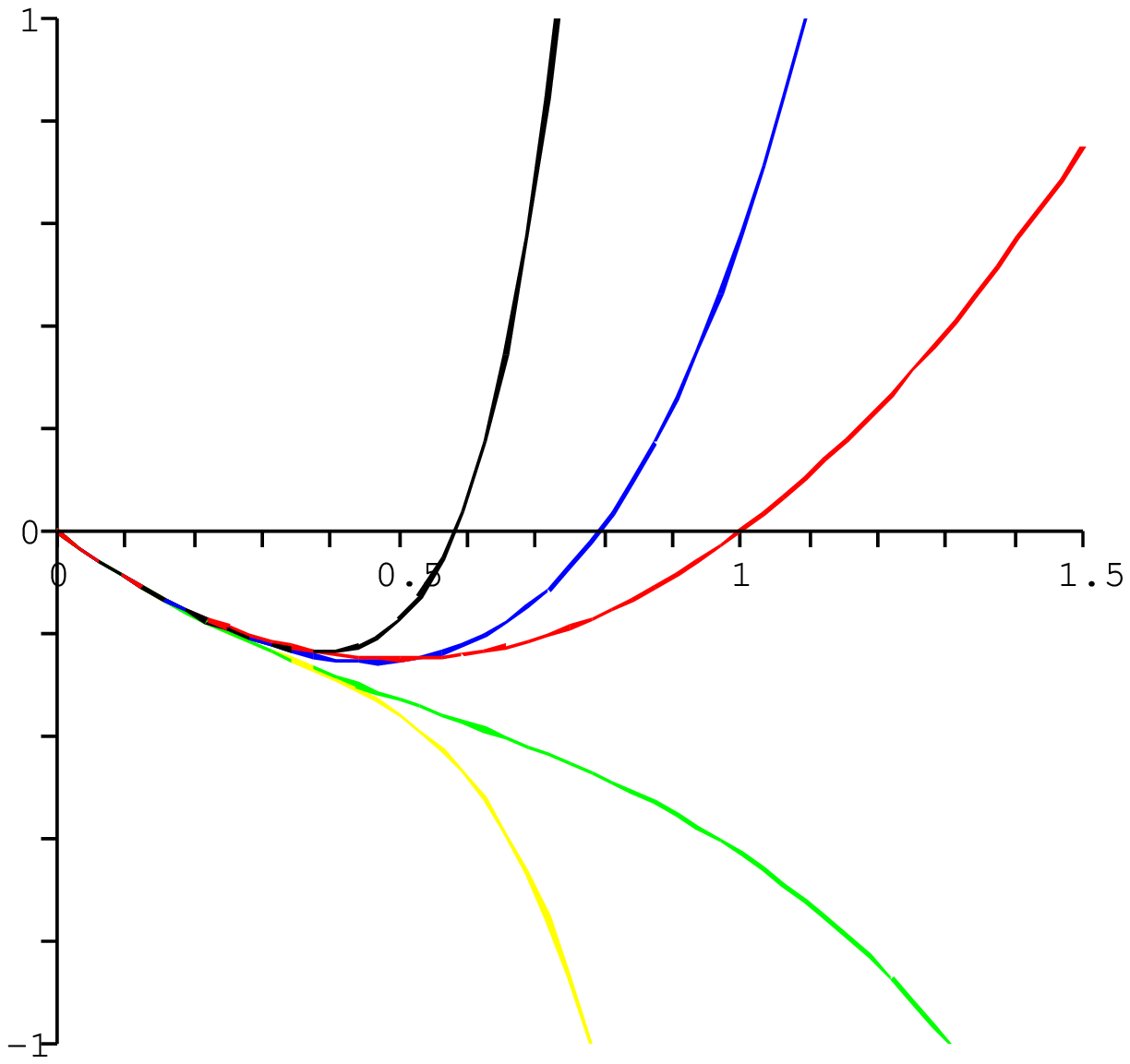}  }}
\put(5,140){\parbox[t]{10mm}{$\beta(u)$}}
\put(170,140){\parbox[t]{10mm}{\small \bf 1}}
\put(128,140){\parbox[t]{10mm}{\small \bf 3}}
\put(95,140){\parbox[t]{10mm}{\small \bf 5}}
 \put(97,35){\parbox[t]{10mm}{\small \bf 4}}
 \put(140,35){\parbox[t]{10mm}{\small \bf 2}}
   \put(180,80){\parbox[t]{10mm}{$u$}}
  \put(85,-5){\parbox[t]{10mm}{(a)}}
 \put(220,20){\parbox[t]{110mm}{\includegraphics[width=
70mm,height=50mm,angle=0]{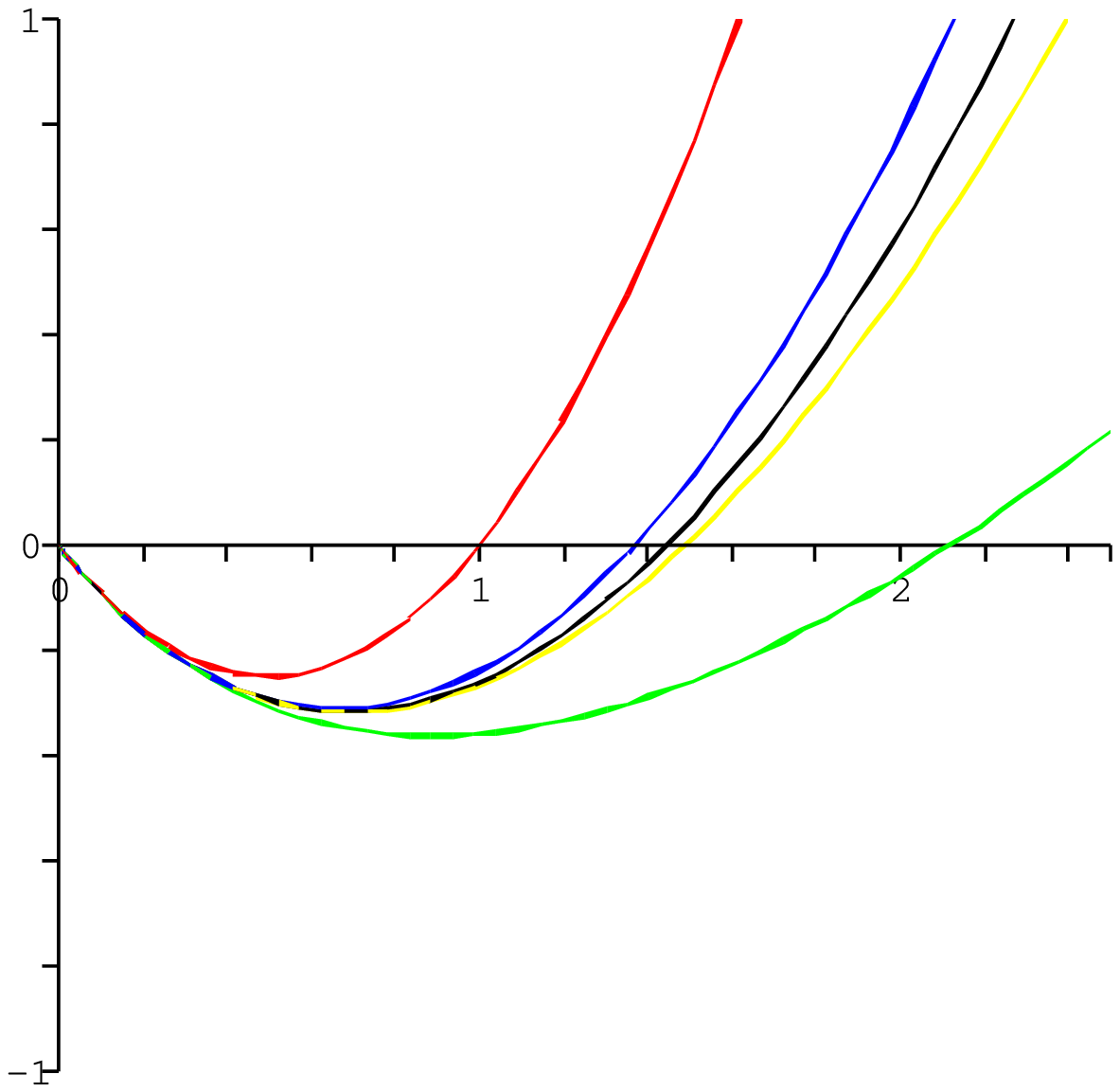} }}
\put(210,140){\parbox[t]{10mm}{$\beta(u)$}}
\put(330,140){\parbox[t]{10mm}{\small \bf 1}}
\put(362,140){\parbox[t]{10mm}{\small \bf 3}}
\put(373,140){\parbox[t]{10mm}{\small \bf 5}}
\put(390,140){\parbox[t]{10mm}{\small \bf 4}}
\put(390,103){\parbox[t]{10mm}{\small \bf 2}}
 \put(400,80){\parbox[t]{10mm}{$u$}}
   \put(325,-5){\parbox[t]{10mm}{(b)}}
 \end{picture}
\caption{\label{fig4} $\beta$-function of the 3d $m=1$ model in
successive perturbation theory orders ranging from 1 to 5 as shown
by the labels in the figures. Left: naive evaluation of the
function (\ref{6.2}). Right: resummation of (\ref{6.2}) taking
into account asymptotic properties of the series. Note a decrease
of the difference between FP coordinates found in successive
orders of the resummed perturbation expansion,
figure~\ref{fig4}{\bf b}.}
\end{figure}
 To give an idea about the RG expansions and their
treatment,  we write down several first terms of the
$\beta$-function (\ref{6.1}) \cite{Kleinert91}:
\begin{equation} \label{6.2}
\beta(u)= -u(\varepsilon -u + 3u^2(3m+14)/(m+8)+ \cdots),
\end{equation}
with $\varepsilon=4-d$. For a quantitative analysis, one can
develop an {\em $\varepsilon$-expansion} looking for solutions of
the FP equation
\begin{equation} \label{6.2a}
\beta(u^*)=0
\end{equation}
in a form of a series $u^*=\sum_iu^{(i)}\varepsilon^i$.
Alternatively, one can solve the non-linear equation (\ref{6.2a})
directly at $d=3$ ($\varepsilon=1$). However, as it is well known
by now, the perturbative  RG expansions have zero radius of
convergence and are asymptotic at best \cite{RGbooks} (cf.
behaviour of the function (\ref{6.2}), figure~\ref{fig4}{\bf a}).
Special procedures of {\em resummation} have been elaborated to
deal with them. We give an example of how one of these procedures
works transforming a divergent series, figure~\ref{fig4}{\bf a},
into a convergent one, figure~\ref{fig4}{\bf b}. In the last
figure, the function (\ref{6.2}) has been resummed by the
Pad\'e-Borel resummation. The procedure consists of several steps.
First, assuming the factorial growth of the coefficients $c_i$ in
the expansion $\beta(u)=\sum_i c_iu^i$ (\ref{6.2}), one constructs
the Borel trransform of the initial function $\beta(u)$ via:
\begin{equation} \label{6.2b}
\beta^{\rm B}(u)= \sum_{i} \frac{c_iu^i}{i!}\,.
\end{equation}
Then, the Borel transform is extrapolated by a Pad\'e approximant
$[K/L](u)$. The last is the ratio of two polynomials of order $K$
and $L$ such that its truncated Taylor expansion is equal to
$\beta^{\rm B}(u)$. The resummed function is then calculated by an
inverse Borel transform of this approximant:
\begin{equation} \label{6.2c}
\beta(u)=\int_0^{\infty}{\rm d}t \exp(-t)[K/L](ut).
\end{equation}
Similar techniques are currently widely used in analyzing the RG
expansions \cite{RGbooks,Holovatch02}. In particular, the
numerical estimates of different physical quantities given in this
chapter (and in Table \ref{tab1} as well) have been obtained using
resummation techniques.

One sees the presence of two FPs in figure~\ref{fig4}{\bf b}: an
unstable Gaussian  FP {\bf G} $u^*=0$ (the slope of the
$\beta$-function is negative, $\partial \beta(u)/\partial u
|_{u^*} < 0$) and a stable one, $u^*\neq 0$. Here, the slope of
the $\beta$-function is positive: starting form any initial
conditions with $u>0$ the solution of the differential equation
(\ref{6.1}) reaches its FP value. This FP corresponds to the
critical point  $T_{\mathrm{c}}$ of the 3d Ising  model. Similar
behaviour of the $\beta$-function is found for other values of
$m$, therefore the FP with $u\neq 0$ for general $m$ is called a $O(m)$-symmetric or
Heisenberg FP {\bf H}. We do not show the procedure of calculating
the critical exponents: the other RG functions being evaluated at
this FP bring about the asymptotic values of the critical
exponents (in particular, those given in Table \ref{tab1}). These
exponents govern criticality of systems of different microscopic
nature, which can be described by the effective Hamiltonian
(\ref{5.14}). It is said that these systems belong to the $O(m)$
{\em universality class}. In the RG picture, different microscopic
origin is reflected in different initial conditions for the flow.
However, the FP location and stability is defined solely by the
global features: dimensionality, symmetry, interaction
type.\footnote{In particular, here we consider systems with the
short-range interaction.} Systems which share the global features
belong to the same universality class.

\subsection{Cubic model}\label{VIb}
\begin{figure}[!h]
\begin{picture}(80,170)
\put(20,10){\parbox[t]{110mm}{\includegraphics[width=
60mm,angle=0]{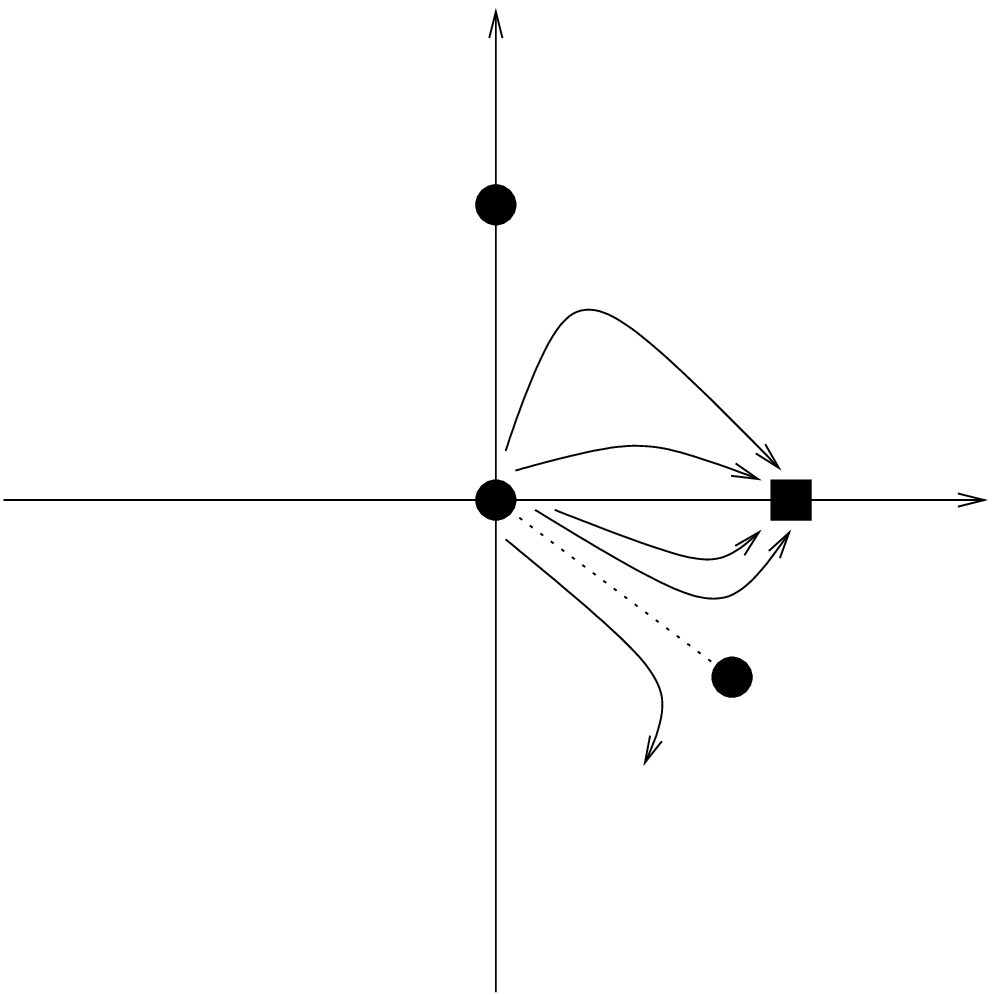}  }}
 \put(95,170){\parbox[t]{10mm}{$v$}}
\put(92,142){\parbox[t]{10mm}{\bf I}}
\put(90,97){\parbox[t]{10mm}{\bf G}}
\put(162,97){\parbox[t]{10mm}{\bf H}}
\put(152,62){\parbox[t]{10mm}{\bf M}}
   \put(180,85){\parbox[t]{10mm}{$u$}}
  \put(100,-5){\parbox[t]{10mm}{(a)}}
 \put(220,10){\parbox[t]{110mm}{\includegraphics[width=
60mm,angle=0]{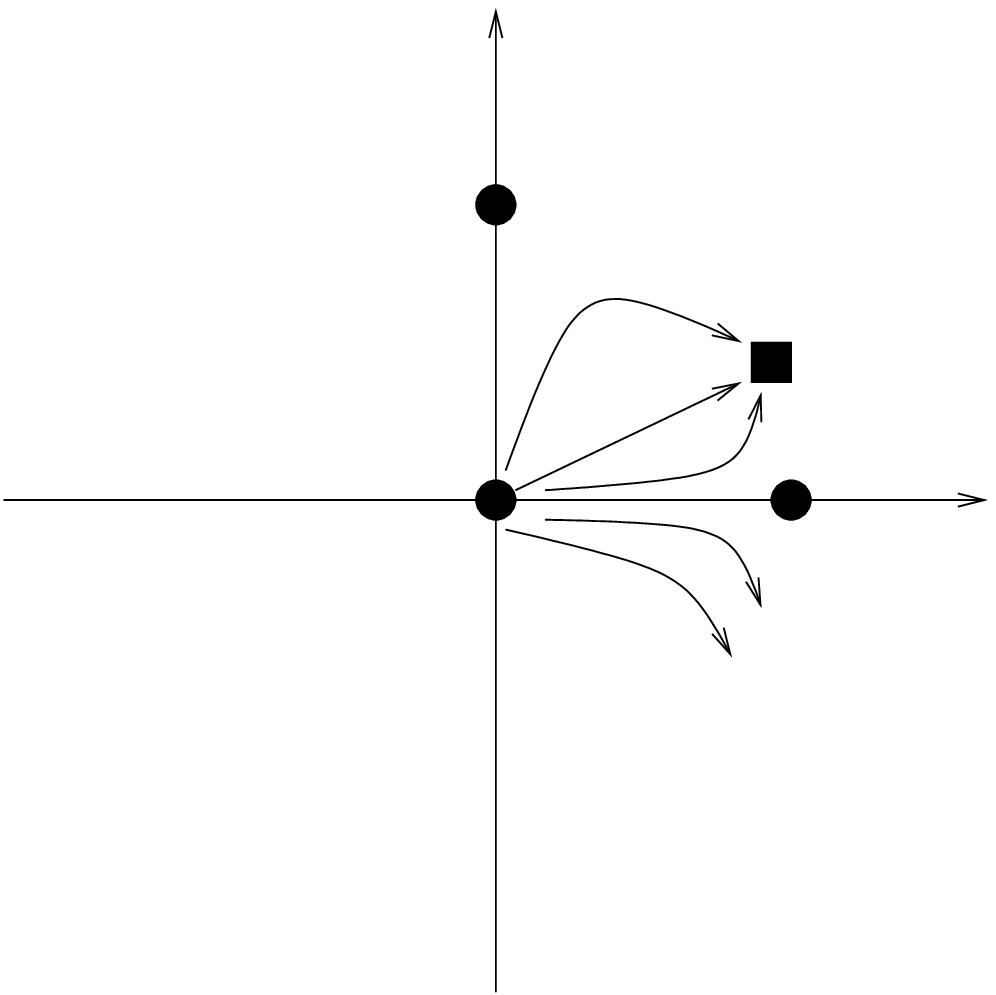} }}
 \put(295,170){\parbox[t]{10mm}{$v$}}
 \put(292,142){\parbox[t]{10mm}{\bf I}}
 \put(290,97){\parbox[t]{10mm}{\bf G}}
 \put(362,97){\parbox[t]{10mm}{\bf H}}
\put(359,117){\parbox[t]{10mm}{\bf M}}
   \put(380,85){\parbox[t]{10mm}{$u$}}
   \put(305,-5){\parbox[t]{10mm}{(b)}}
 \end{picture}
\caption{\label{fig5} FPs and RG flows of the cubic model.
Unstable FPs are shown by discs, stable FPs are shown by squares.
(a): $m<m_{\mathrm{c}}^{\rm cub}$, the Heisenberg FP {\bf H} is
stable. (b): $m>m_{\mathrm{c}}^{\rm cub}$, the mixed FP {\bf M} is
stable. The dotted line in figure~\ref{fig5}{\bf a} shows a
separatrix: the RG flows which start below this line do not reach
{\bf H}. Note that for $m>m_{\mathrm{c}}^{\rm cub}$ the $u$-axis
is a separatrix: all flows that start from the initial conditions
with $v<0$ do not reach any FP (run-away solutions).}
\end{figure}

The main difference of the effective Hamiltonian of the cubic
model (\ref{5.14b}) as compared to that of the $m$-vector model
(\ref{5.14}) is that it contains one more coupling $v$ of
different symmetry. Therefore, two $\beta$-functions describe the
RG flow:
\begin{equation} \label{6.3}
 \frac{\rd u}{\rd\ln \ell}=\beta_u(u,v), \hspace{3em}
 \frac{\rd v}{\rd\ln \ell}=\beta_v(u,v).
\end{equation}
Stability  of a FP is now defined by the stability matrix composed
of the $\beta$-function derivatives $\partial \beta_{u,v}/\partial
\{u,v\}$. The FP picture, which arises form an analysis of the
$\beta$-functions at $d=3$ \cite{Aharony76} is schematically shown
in figure~\ref{fig5}. Four FPs are obtained: unstable Gaussian
{\bf G} ($u=v=0$) and Ising {\bf I} ($u=0, v\neq 0$) as well as
Heisenberg {\bf H} ($u\neq 0, v= 0$) and mixed {\bf M} ($u\neq0,
v\neq0$). The stability of the two last FPs depends on the value
of $m$.

Two different regimes for the RG flows are observed. For small
$m<m_{\mathrm{c}}^{\rm cub}$ the FP {\bf H} is stable. At this FP
the system does not feel a presence of the cubic coupling, $v=0$.
Therefore, for $m<m_{\mathrm{c}}^{\rm cub}$ the cubic model
belongs to the $O(m)$ {\em universality class}: its exponents
coincide with the exponents of the $m$-vector model, table
\ref{tab1}. However, with an increase of $m$ the FP {\bf M}
approaches {\bf H} and at $m=m_{\mathrm{c}}^{\rm cub}$ both FPs
coincide: a {\em crossover} to the new regime occurs. For
$m>m_{\mathrm{c}}^{\rm cub}$ the FP {\bf M} becomes stable and
governs the critical properties of the cubic model in the new
universality class. The {\em marginal dimension} value is slightly
less than three: $m_{\mathrm{c}}^{\rm cub}=2.862(5)$
\cite{Folk00}. From this estimate it follows in particular, that a
cubic Heisenberg ($m=3$) magnet does not belong to the $O(3)$
universality class. Its critical exponent being estimated in the
fixed point {\bf M} read \cite{Carmona00}:
\begin{equation} \label{6.4}
 \gamma= 1.390(12), \quad
 \nu= 0.706(6), \quad
 \eta= 0.0333(26) ,\quad
 \beta= 0.364(15), \quad
 \alpha= -0.118(18).
\end{equation}
Numerically, these values are close to their counterparts for the
$m$-vector model (cf. Table \ref{tab1}). However the principal
difference arises from the above analysis: as one can easily check
solving the system of differential equations (\ref{6.3}) for
$m>m_{\mathrm{c}}^{\rm cub}$ the RG flows with $v<0$ cannot reach
the stable FP (an abscissa $v=0$ serves as a separatrix for the
flows). As it follows from sections \ref{IIIb}, \ref{Vb} negative
$v$ corresponds to ordering along diagonals of $m$-dimensional
hypercube. Therefore, the RG analysis results in a statement that
such ordering cannot occur via a 2nd order phase transition:
ferromagnetic crystals with three easy axes should undergo a 1st
order phase transition. It is worth noting here, that whereas the
presence of stable and reachable FP brings about the 2nd order
phase transition, its absence signalling only that the 2nd order
phase transition does not occur. The nature of the low-temperature
phase and  the scenario of how it is attained remains to be
checked by other methods.

\subsection{Random-site dilution}\label{VIc}

\begin{figure}[!h]
\begin{picture}(80,170)
\put(20,10){\parbox[t]{110mm}{\includegraphics[width=
60mm,angle=0]{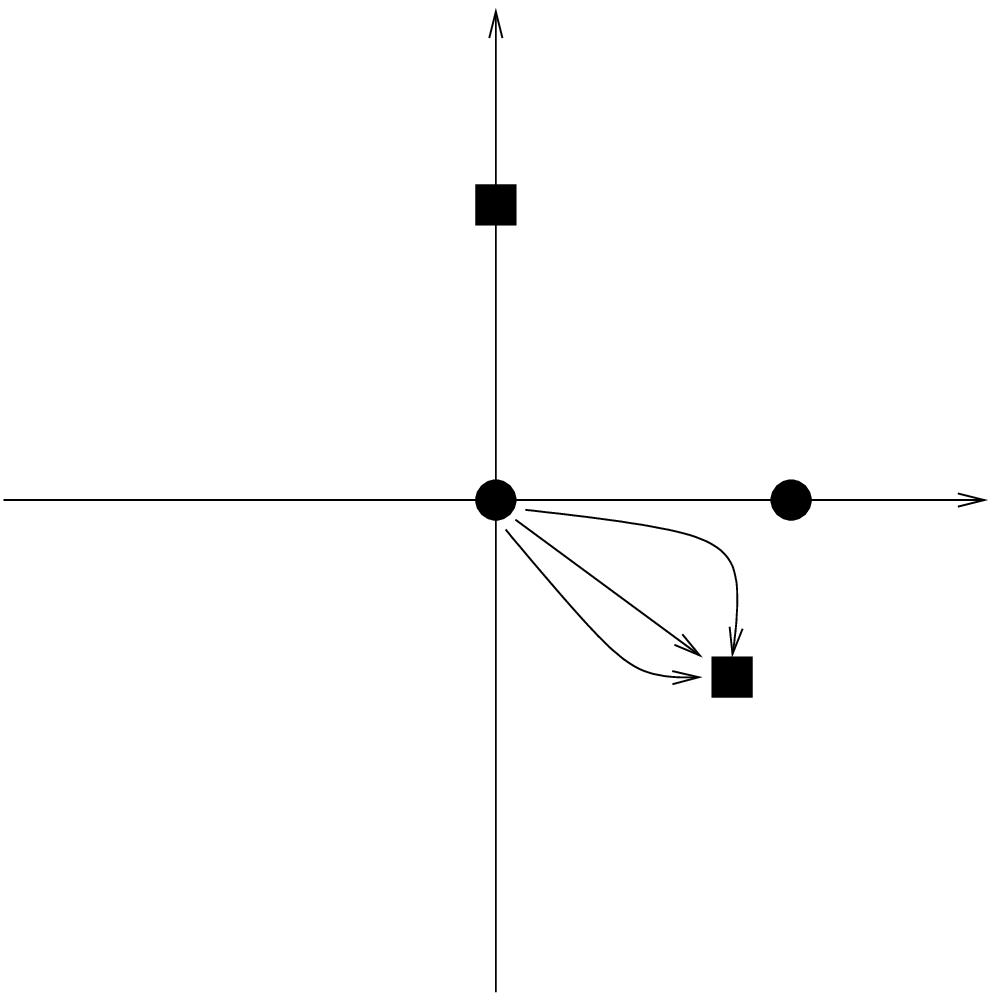}  }}
 \put(95,170){\parbox[t]{10mm}{$v$}}
\put(92,145){\parbox[t]{10mm}{\bf P}}
\put(90,97){\parbox[t]{10mm}{\bf G}}
\put(162,97){\parbox[t]{10mm}{\bf H}}
\put(152,62){\parbox[t]{10mm}{\bf M}}
   \put(180,85){\parbox[t]{10mm}{$u$}}
  \put(100,-5){\parbox[t]{10mm}{(a)}}
 \put(220,10){\parbox[t]{110mm}{\includegraphics[width=
60mm,angle=0]{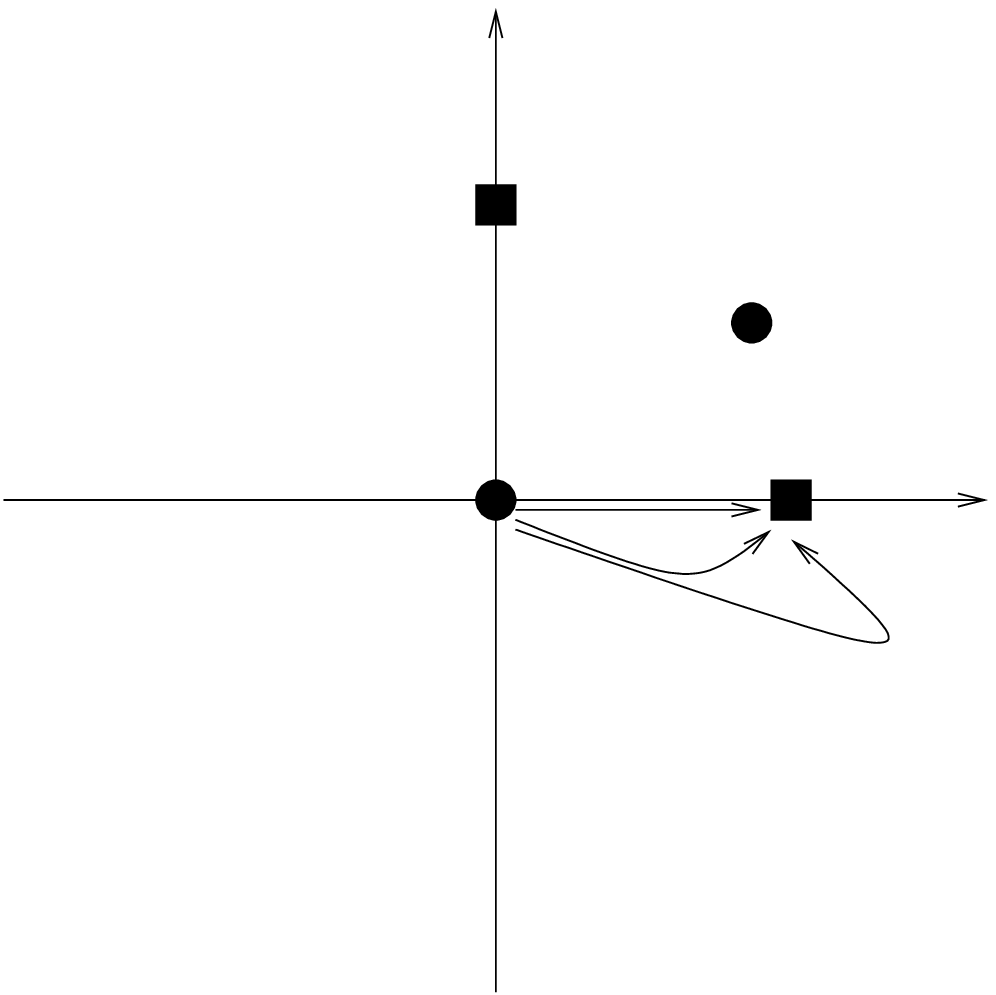} }}
 \put(295,170){\parbox[t]{10mm}{$v$}}
 \put(292,145){\parbox[t]{10mm}{\bf P}}
 \put(290,97){\parbox[t]{10mm}{\bf G}}
 \put(362,97){\parbox[t]{10mm}{\bf H}}
\put(356,123){\parbox[t]{10mm}{\bf M}}
 \put(329,88){\parbox[t]{10mm}{\tiny \bf 1}}
 \put(340,85){\parbox[t]{10mm}{\tiny \bf 2}}
 \put(359,67){\parbox[t]{10mm}{\tiny \bf 3}}
   \put(380,85){\parbox[t]{10mm}{$u$}}
   \put(305,-5){\parbox[t]{10mm}{(b)}}
 \end{picture}
\caption{\label{fig6} FPs and RG flows of the weakly diluted
quenched $m$-vector model. Unstable FPs are shown by discs, stable
FPs are shown by squares. (a): $m<m_{\mathrm{c}}^{\rm dil}$, the
mixed FP {\bf M} is stable. (b): $m>m_{\mathrm{c}}^{\rm dil}$, the
Heisenberg FP {\bf H} is stable. Flows 1, 2, 3 of the
figure~\ref{fig6}{\bf b} are further treated in
figure~\ref{fig7}{\bf b}.}
\end{figure}

Formally, an analysis of the weakly diluted quenched $m$-vector
model resembles those we discussed in the former subsection
\ref{VIb}. Indeed, both effective Hamiltonians (\ref{5.14b}),
(\ref{5.25}) contain two couplings of different symmetry and give
rise to the already familiar FP picture. Moreover, the effective
Hamiltonian  (\ref{5.25}) at $m=1$ coincides with
(\ref{5.14b}).\footnote{One can check it by further substitution
$n\rightarrow m$, $\{u_0,v_0\} \rightarrow \{v_0,u_0\}$.} However,
our goal is to analyze it in the replica limit $n=0$. Note, that
now the physically meaningful values of couplings are $u>0$,
$v<0$, see section \ref{Vc}. The FP picture and the RG flows at
$d=3$ are shown in the figure~\ref{fig6} \cite{Folk03}. Besides
the familiar FPs {\bf G}, {\bf H}, and {\bf M} a polymer FP {\bf
P} is present. It is stable and corresponds to the $O(m=0)$
universality class, however it is never reached from the initial
conditions $u>0$, $v<0$.

Again, the new marginal dimension $m_{\mathrm{c}}^{\rm dil}$
governs the crossover between the new  and the $O(m)$ universality
classes: FP {\bf M} is stable for $m<m_{\mathrm{c}}^{\rm dil}$. A
search for the value of $m_{\mathrm{c}}^{\rm dil}$ shows, that at
$m=m_{\mathrm{c}}^{\rm dil}$ the heat capacity critical exponent
of an undiluted system changes its sign:
$\alpha(m_{\mathrm{c}}^{\rm dil})=0$. In this way one recovers the
Harris criterion \cite{Harris74}, section \ref{IIIc1}, translated
into the RG ``language''. The numerical value of
$m_{\mathrm{c}}^{\rm dil}$ being slightly less than two,
$m_{\mathrm{c}}^{\rm dil}=1.912(4)$ \cite{Holovatch01}, only the
Ising model ($m=1$) changes its exponents upon dilution. Indeed,
the numerical values of the exponents read~\cite{Pelissetto00}:
\begin{equation} \label{6.5}
 \gamma= 1.330(17), \quad
 \nu= 0.678(10), \quad
 \eta= 0.030(3),\quad
 \beta= 0.349(5),\quad
 \alpha= -0.034(30)
\end{equation}
and differ essentially from those of the 3d Ising model (table
\ref{tab1}).

\begin{figure}[ht]
\begin{picture}(80,170)
\put(20,10){\parbox[t]{110mm}{\includegraphics[width=
60mm,angle=0]{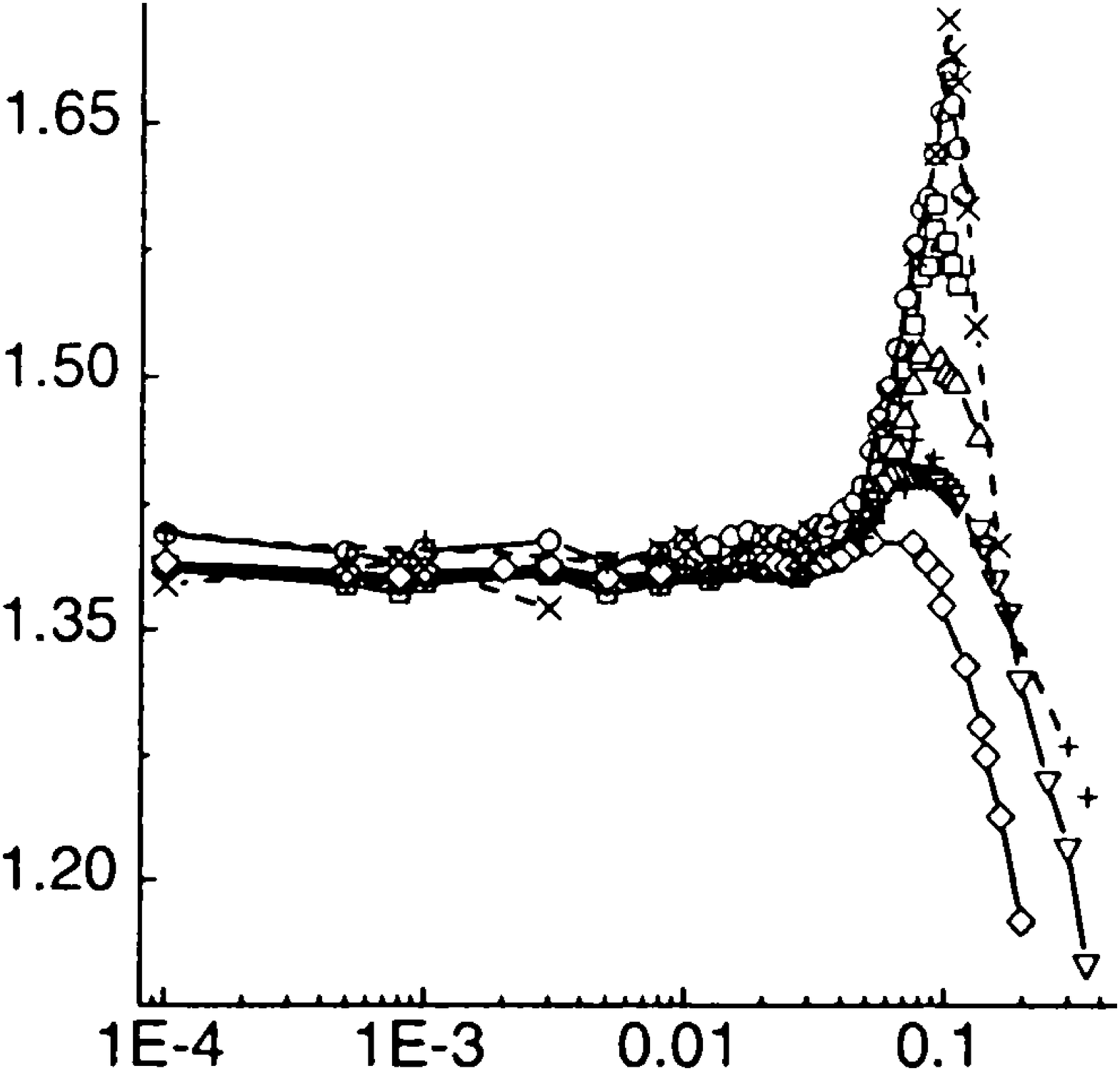}  }}
 \put(25,170){\parbox[t]{10mm}{$\gamma^{\rm eff}$}}
   \put(195,17){\parbox[t]{10mm}{$\tau$}}
  \put(100,-5){\parbox[t]{10mm}{(a)}}
 \put(220,4){\parbox[t]{110mm}{\includegraphics[height=60mm,width=
60mm,angle=0]{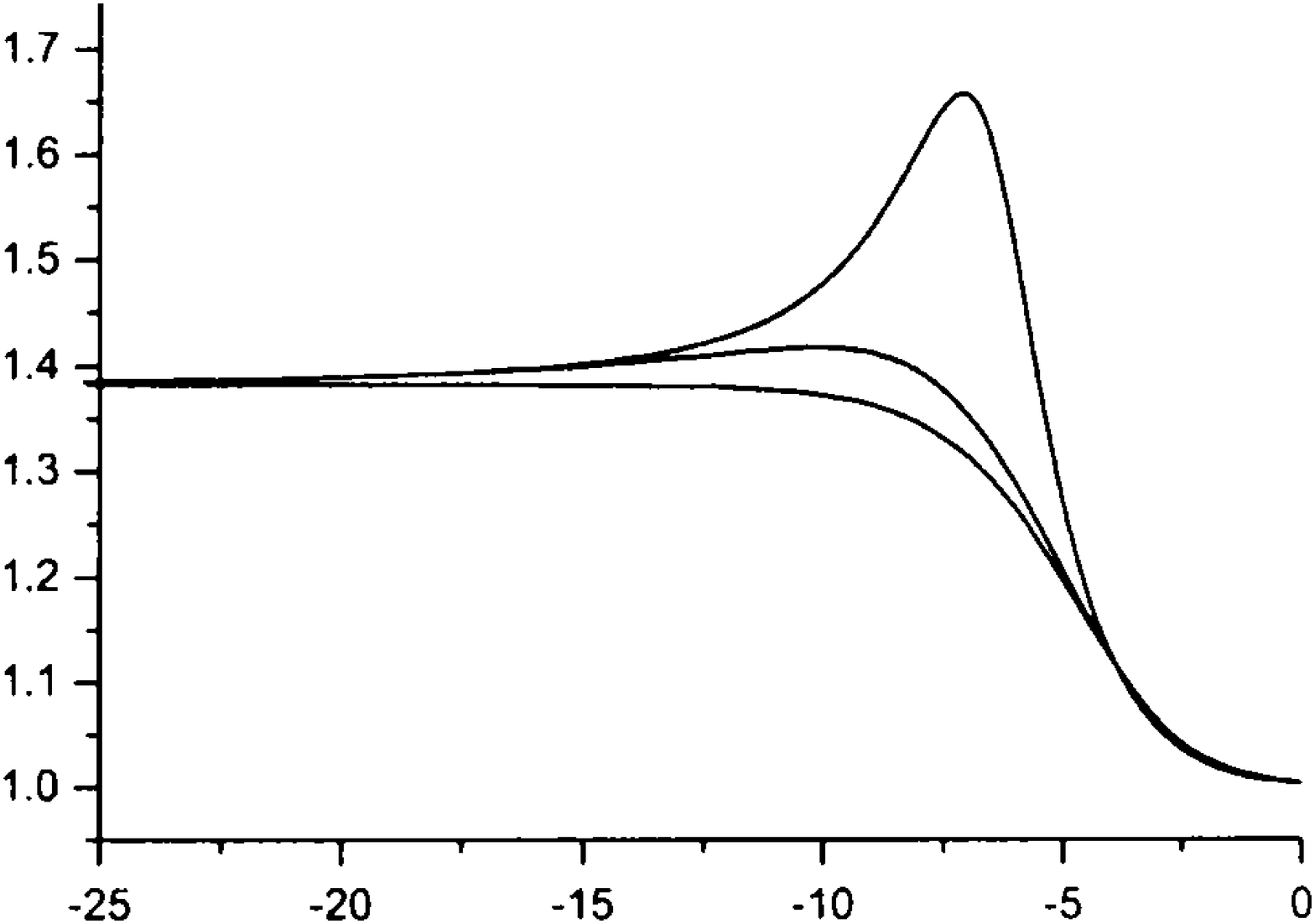} }}
\put(205,170){\parbox[t]{10mm}{$\gamma^{\rm eff}$}}
   \put(390,17){\parbox[t]{10mm}{$\ln \ell$}}
   \put(329,91){\parbox[t]{10mm}{\tiny \bf 1}}
 \put(329,115){\parbox[t]{10mm}{\tiny \bf 2}}
 \put(329,140){\parbox[t]{10mm}{\tiny \bf 3}}
   \put(305,-5){\parbox[t]{10mm}{(b)}}
 \end{picture}
\caption{\label{fig7} Effective critical exponent $\gamma^{\rm eff
}$ for the magnetic susceptibility of the weakly diluted quenched
Heisenberg magnet. (a): as a function of the distance to the Curie
point $\tau=(T-T_{\mathrm{c}})/T_{\mathrm{c}}$ in experimental
measurements for a-${\rm Fe_{86}Mn_4Zr_{10}}$ amorphous alloy
\cite{Perumal03}. (b): as a function of the RG flow parameter in
theoretical RG calculations for the diluted $m=3$ model
\cite{Dudka03}. Different curves correspond to different amount of
disorder. Note that in the asymptotics ($\tau\rightarrow 0$ or
$\ell \rightarrow 0$) the exponent attains its universal value.}
\end{figure}

However, both in the experiments and in the MC simulations one
deals with the system not yet in an asymptotic region, where
exponents do not attain their FP values and the {\em effective
exponents}  are found. Being non-universal, they can be calculated
in the RG treatment as functions of the flow-dependent couplings.
In figure~\ref{fig7}, we show an effective critical exponent
$\gamma_{\rm eff}$ measured recently for the ac susceptibility of
the a-${\rm Fe_{86}Mn_4Zr_{10}}$ amorphous alloy \cite{Perumal03}
and compare them with the theoretical RG calculations of the
effective exponents of the weakly diluted quenched $m=3$ model.
The exponent was calculated along different RG flows labeled by
numbers in figure~\ref{fig6}{\bf b}. This calculation serves an
example how the non-asymptotic effects may be taken into account
in the RG analysis. Although direct correspondence between the
temperature distance to the critical point $\tau$ and the RG flow
parameter $\ell$ is problematic, the RG serves as a useful tool of
accompanying the studies of effective critical behaviour.

\subsection{Random anisotropy}\label{VId}

\begin{figure}[h]
\includegraphics[width=0.48\textwidth]{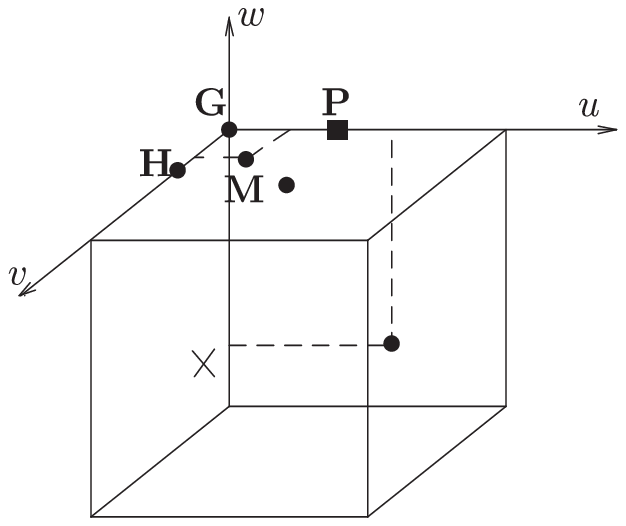}%
\hfill%
\includegraphics[width=0.48\textwidth]{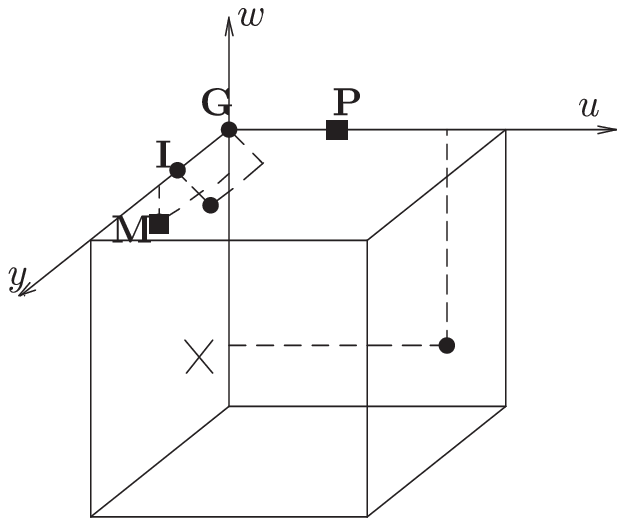}%
\\%
\parbox[t]{0.48\textwidth}{%
\centerline{(a)}%
}%
\hfill%
\parbox[t]{0.48\textwidth}{%
\centerline{(b)}%
}%
\caption{Fixed points of the random anisotropy model. The filled
boxes show the stable fixed points, the cross denotes typical
initial values of couplings. (a): isotropic local random axis
distribution. The stable FP {\bf P} cannot be reached by the RG
flow which starts from the region shown by a cross in the figure.
(b): cubic local random axis distribution. The stable FP {\bf P}
cannot be reached, but the random Ising FP {\bf M} is stable and
reachable for the RG flow.} \label{fig8}
\end{figure}

On this example we shall show how the RG predicts two different
phenomena occuring in the random anisotropy magnets. It appears
that the type of local random axis distribution crucially effects
an origin of the low-temperature phase in random anisotropy
systems \cite{ramreviews}. First, we consider the results obtained
for an isotropic distribution (\ref{3.7}), which leads to the
effective Hamiltonian (\ref{5.29}) \cite{Aharony75}. It contains
three couplings, $u,v,w$, hence three $\beta$-functions define the
RG flows. Solving the FP equations one arrives at the FP picture
shown in figure~\ref{fig8}{\bf a} \cite{Aharony75,Dudka05}.

Let us recall (cf. section \ref{Vd}), that physically meaningful
values of couplings are $u>0,\linebreak v>0,w<0$. Therefore only
the FPs located in this region are shown in the figure. However,
there is another condition found for the ratio of couplings:
$w_0/u_0=-m$. The region of typical initial conditions to study
the RG flow is shown in figure~\ref{fig8} by a cross. The only
stable FP found, a polymer FP {\bf P} is not reachable from the
initial conditions. The {\em run-away} solutions of the RG
equations bring about an absence of a 2nd order phase transition.

A different picture is obtained for a cubic random axis
distribution (\ref{3.8}) \cite{Dudka05}. Here, the effective
Hamiltonian (\ref{5.30}) contains four couplings of different
symmetry, $u$, $v$, $w$, $y$. The physical initial values for the
couplings lay in the region (section \ref{Vd}):
$u>0,v>0,w<0,w/u=-m$. A typical FP picture is shown in
figure~\ref{fig8}{\bf b} for $v=0$. Similar to the former case of
isotropic random axis distribution, the stable FP {\bf P} cannot
be reached. However, one more stable FP {\bf M} is present. It is
reachable for the RG flow that starts from the initial conditions
marked by a cross in the figure. This FP is a FP of the
random-site Ising model (subsection \ref{VIc}) for any value of
$m$. It means that the ferromagnetic 2nd order phase transition in
the $m$-vector magnet with the cubic random axis distribution
belongs to the universality class of the random-site Ising model
and is governed in asymptotics by the exponents (\ref{6.5}).

\subsection{Stacked triangular antiferromagnet}\label{VIe}

\begin{figure}[ht]
\begin{picture}(80,390)
\put(20,215){\parbox[t]{110mm}{\includegraphics[width=
60mm,angle=0]{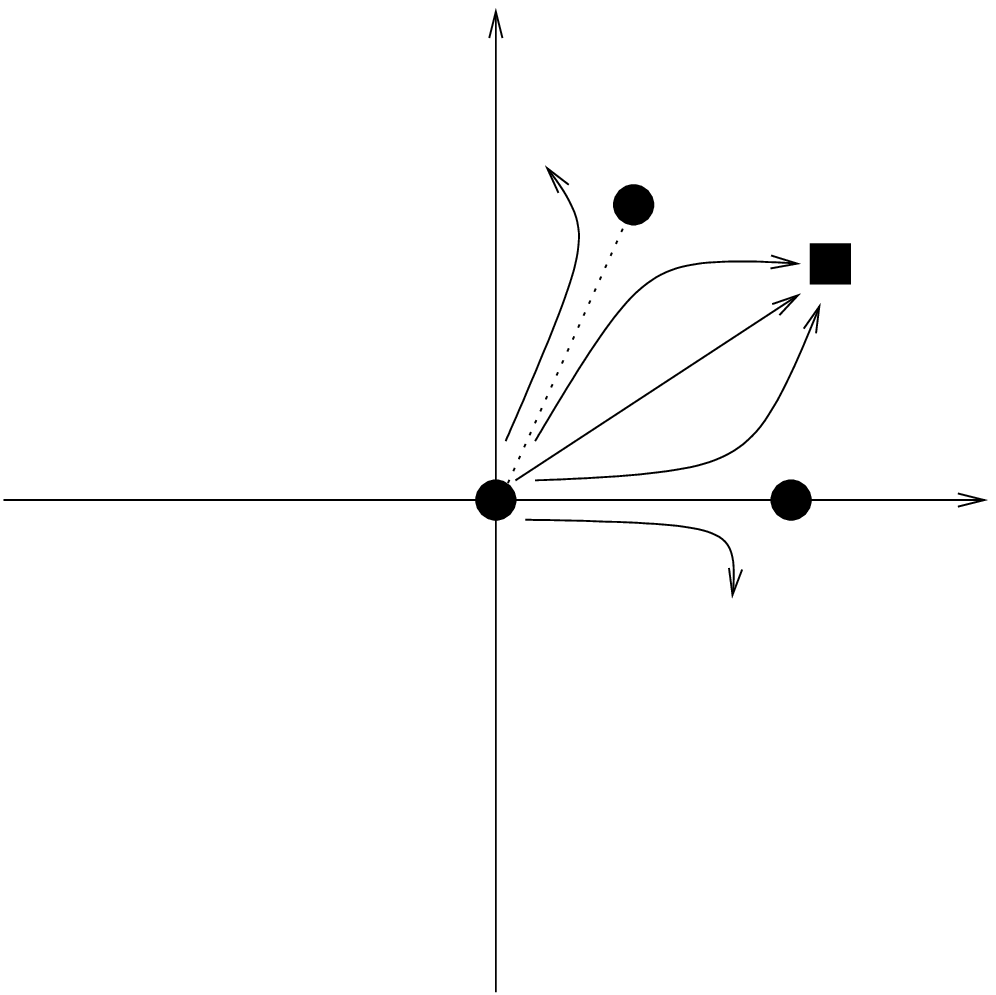}  }}
 \put(85,200){\parbox[t]{30mm}{${\bf m>m_3^{\rm chir}}$}}
 \put(95,375){\parbox[t]{10mm}{$v$}}
   \put(180,290){\parbox[t]{10mm}{$u$}}
   \put(90,302){\parbox[t]{10mm}{\bf G}}
\put(162,302){\parbox[t]{10mm}{\bf H}}
 \put(130,355){\parbox[t]{10mm}{{\bf C}$_-$}}
\put(167,330){\parbox[t]{10mm}{{\bf C}$_+$}}
  \put(220,215){\parbox[t]{110mm}{\includegraphics[width=
60mm,angle=0]{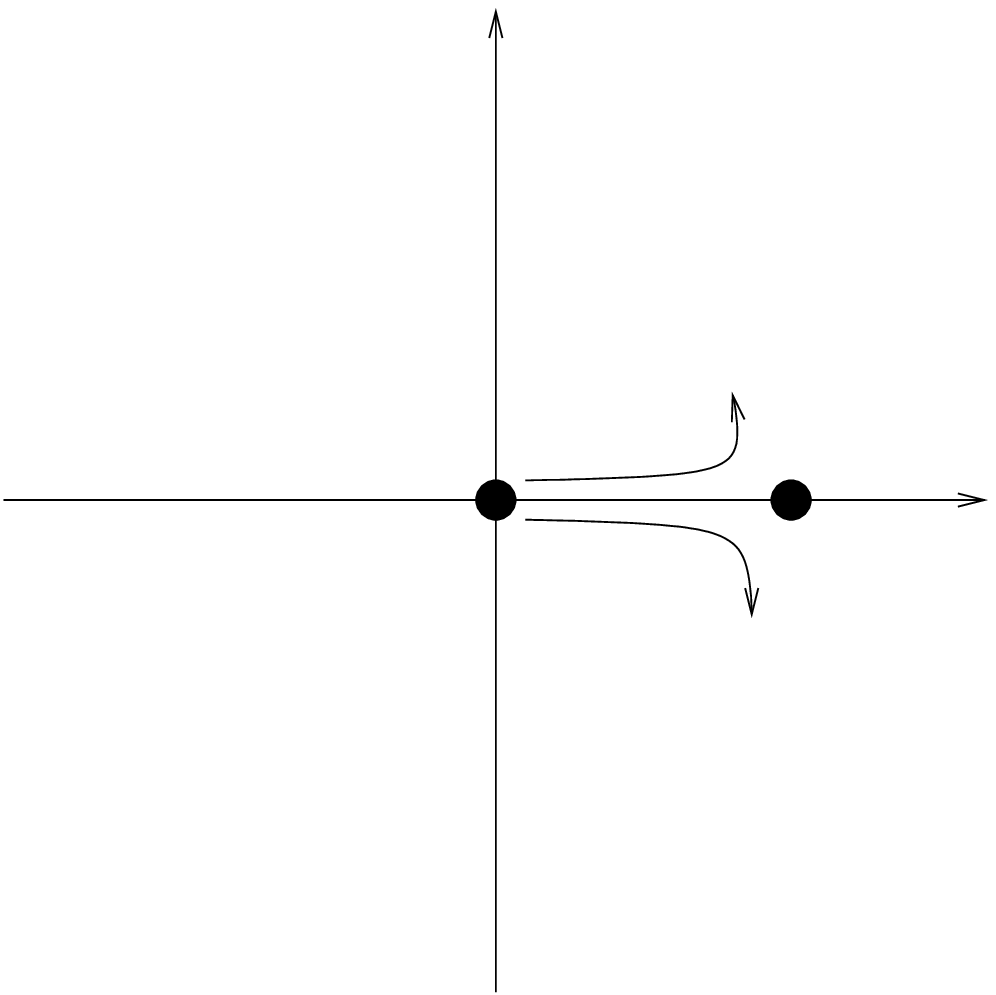} }}
 \put(295,375){\parbox[t]{10mm}{$v$}}
  \put(380,290){\parbox[t]{10mm}{$u$}}
  \put(290,302){\parbox[t]{10mm}{\bf G}}
 \put(362,302){\parbox[t]{10mm}{\bf H}}
 \put(260,200){\parbox[t]{50mm}{${\bf m_3^{\rm chir}>m>m_2^{\rm chir}}$}}
\put(20,15){\parbox[t]{110mm}{\includegraphics[width=
60mm,angle=0]{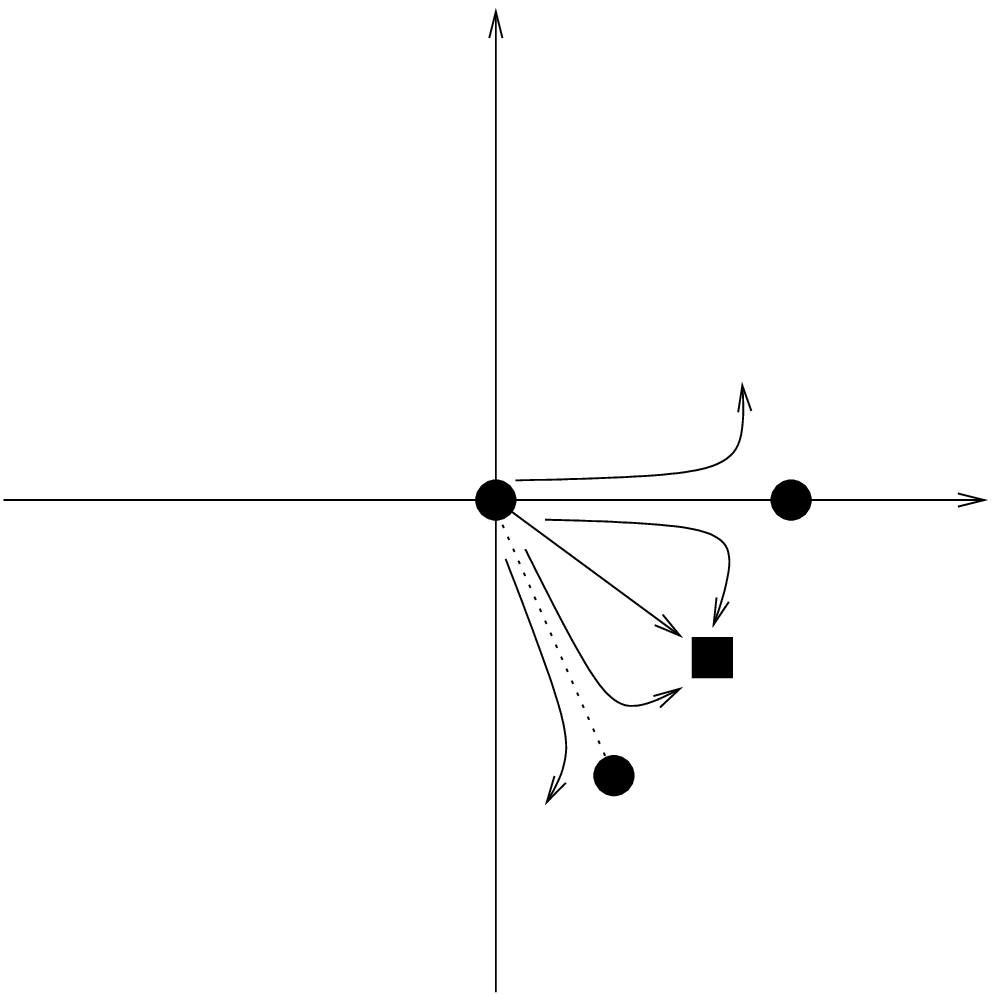}  }}
 \put(95,175){\parbox[t]{10mm}{$v$}}
\put(90,102){\parbox[t]{10mm}{\bf G}}
\put(162,102){\parbox[t]{10mm}{\bf H}}
\put(152,67){\parbox[t]{10mm}{{\bf S}$_+$}}
\put(130,45){\parbox[t]{10mm}{{\bf S}$_-$}}
   \put(180,90){\parbox[t]{10mm}{$u$}}
  \put(62,0){\parbox[t]{50mm}{${\bf m_2^{\rm chir}>m>m_1^{\rm chir}}$}}
 \put(220,15){\parbox[t]{110mm}{\includegraphics[width=
60mm,angle=0]{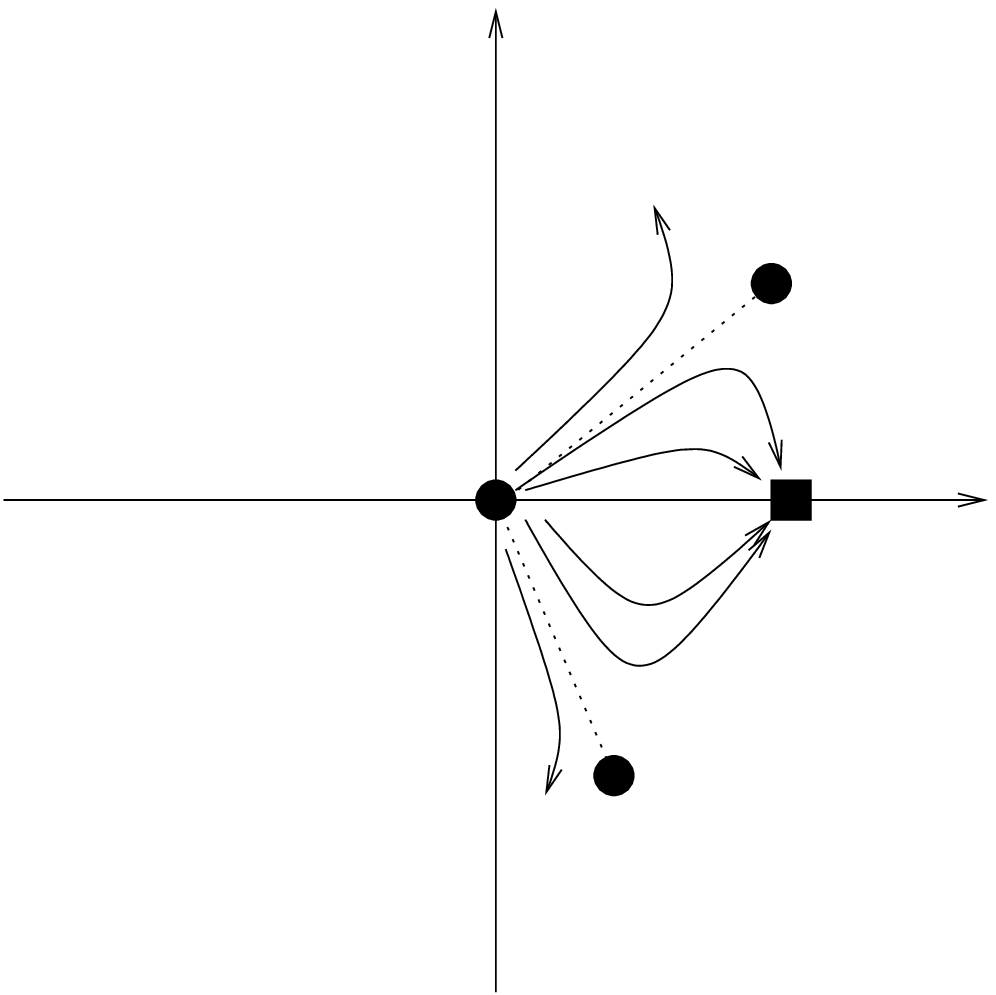} }}
 \put(295,175){\parbox[t]{10mm}{$v$}}
  \put(290,102){\parbox[t]{10mm}{\bf G}}
 \put(362,102){\parbox[t]{10mm}{\bf H}}
\put(356,128){\parbox[t]{10mm}{{\bf S}$_+$}}
\put(330,45){\parbox[t]{10mm}{{\bf S}$_-$}}
    \put(380,90){\parbox[t]{10mm}{$u$}}
   \put(285,0){\parbox[t]{30mm}{${\bf m<m_1^{\rm chir}}$}}
 \end{picture}
\caption{\label{fig9} FPs and RG flows of the stacked triangular
antiferromagnet model. Unstable FPs are shown by discs, stable FPs
are shown by squares. Three marginal dimensions $m_1^{\rm chir}$,
$m_2^{\rm chir}$, $m_3^{\rm chir}$ govern the FP picture.}
\end{figure}

Again, as in the former subsections \ref{VIa}--\ref{VId}, the RG
answer about a possibility of a 2nd order phase transition in the
3d stacked triangular antiferromagnet would be a presence of a
stable accessible FP for the couplings $u,v$ of the effective
Hamiltonian (\ref{5.32}). The model possesses a rather complicated
FP structure, sketched in figure~\ref{fig9}.

FP picture changes with $m$ and one finds three marginal
dimensions $m_i^{\rm chir}$ that govern its topology. For large
$m>m_3^{\rm chir}$ the stable chiral FP {\bf C}$_+$ is present and
it can be reached from the initial values of the couplings
$u,v\geqslant  0$. Therefore, the phase transition to the
non-collinear chiral state is of the second order. At $m=m_3^{\rm
chir}$, {\bf C}$_+$ merges with the anti-chiral FP {\bf C}$_-$ and
disappears, only one unstable Heisenberg FP {\bf H} is found for
$m_3^{\rm chir}>m>m_2^{\rm chir}$. With further decrease of $m$,
two more FPs appear at $m=m_2^{\rm chir}$,  {\bf S}$_-$ and {\bf
S}$_+$. The last one is stable, however it describes the
sinusoidal phase, which occurs for $v<0$. Finally, at $m=m_1^{\rm
chir}$  this FP merges with {\bf H} and looses its stability with
further increase of $m$. It follows from this analysis, that the
2nd order phase transition into non-collinear phase can occur only
in the magnets with $m>m_3^{\rm chir}$.

In principle, the above described FP picture has been known since
the first RG studies of the problem \cite{Kawamura88}. However,
reliable numerical estimates for marginal dimensions $m_i^{\rm
chir}$ have been obtained only recently. The question of interest
is, what is the order of the phase transition at  $m=2;3$ when the
model has its physical realization? The estimates: $m_3^{\rm
chir}=6.23(21)$, $m_2^{\rm chir}=1.99(4)$, $m_1^{\rm
chir}=1.43(2)$ \cite{Holovatch04} clearly rule out the possibility
of a 2nd order phase transition for $m=2;3$: the FP picture is
shown in the second figure of figure~\ref{fig9} and no accessible
FP is found.

Note, however, certain controversy in the RG studies performed so
far: whereas the above FP picture is confirmed by the perturbative
RG expansions and the non-perturbative RG approach
\cite{Delamotte04,Holovatch04}, an analysis of the FP equations by
direct solution of the non-linear FP equations for the resummed
$\beta$-functions brings about a presence of the stable reachable
FP for $m=2;3$ \cite{Pelissetto01}. The last is associated with
the critical point of the 2nd order phase transition.

\section{Conclusions and outlook} \label{VII}

How do the changes in structure effect the critical behaviour of
the matter? We tried to give an answer to this question taking a
3d $m$-vector model as an ideal reference system and showing what
will happen to it under an effect of different non-idealities. The
examples considered include anisotropy, structural disorder,
frustrations: the features  one often encounters dealing with
realistic condensed matter objects. The response of a system to
such non-idealities appears to be very different, ranging from an
insensitivity (the cubic $m=2$ magnet remains in the $O(2)$
universality class), through softening (the heat capacity of the
random site $m=1$ magnet does not diverge) to disappearance of a
2nd order phase transition  (isotropically distributed local
random axis destroys long-range order). These various asymptotic
features are accompanied by a complicated non-asymptotic effective
critical behaviour.

It is astonishing that all this bunch of phenomena can be
explained and accurately described within one theoretical
framework, the RG approach. An application of basic RG notions of
flows, fixed points and their accessibility, marginal dimensions
and crossovers supported by an elaborate machinery to perform and
analyze the RG transformation resulted in a coherent picture of
phenomena in the vicinity of a critical point. A lot remains to be
cleared up in this picture. Maybe a participant of the school or a
reader of these lectures will decide to make his or her
contribution? Good luck!

\section*{Acknowledgements} \label{A}

I am grateful to  Bertrand Berche, Arnaldo Donoso,  and Ricardo
Paredes for the invitation to lecture at the Spring school on
Foundations of statistical and mesoscopic physics (Mochima,
Venezuela, June 20th--24th 2005) and to all participants of the
school for the wonderful atmosphere created there. Bertrand Berche
is further acknowledged for his encouragement, advice during
preparation of the manuscript and, last but not least, for his
stories about the Cagniard de la Tour state, the critical state
discovered as early as in 1822 \cite{history}! I thank my
colleagues Viktoria Blavats'ka, Bertrand Delamotte, Maxym Dudka,
Christian von Ferber, Reinhard Folk, Dmytro Ivaneiko, Taras
Yavors'kii  -- some of the results mentioned in the last part of
these lectures are due to our common work.

This work was supported by Austrian Fonds zur F\"orderung der
wi\-ssen\-schaft\-li\-chen Forschung under Project No. P16574.

\section*{Questions and answers}
\begin{itemize}
\item[${\cal Q}$] {\em (Alexander L\'opez)}: How do the signs of the
couplings $u$ and $v$ in the effective Hamiltonian (\ref{5.14b})
determine a type of the low-temperature ordering?

\item[${\cal A}$] Neglecting fluctuations (taking a function
$\phi(r)$ to be just a variable $\phi$) you can think about the
effective Hamiltonian as of the Landau free energy. Now, let the
reference system display a 2nd order phase transition. This means
that $u>0$. Minimizing Landau free energy and looking for the
spontaneous magnetization one finds at $T<T_{\mathrm{c}}$ two
different non-trivial solutions: $\vec{\phi}=(\phi/\sqrt{m},
\dots, \phi/\sqrt{m})$, it exists at $v<0$, and $\vec{\phi}=(\phi,
0, \dots, 0)$ at $v>0$. They correspond to two types of ordering:
along the diagonals or along the edges of a $m$-dimensional
hypercube. For $m=3$, these are directions [111] and [100]
correspondingly.

\item[${\cal Q}$] {\em (Bertrand Berche)}: What are experimental
realizations of the cubic model?

\item[${\cal A}$] I have already mentioned ferromagnetic crystals. Besides,
at $m=3$ the model describes a ferroelectric phase transition
which occurs in ${\rm SrTiO_3}$ at 105 K ( Cowley~R.A.,   Bruce~A.D.,  J.~Phys. C, 1973, {\bf 6}, L191). Moreover, since the model
provides an example of a system with an arbitrary weak first-order
phase transition, it is also used  as a testing ground to describe
an elecroweak transition in the early Universe (P. Arnold, S
Sharpe, L. Yaffe, Y. Zhang, Phys. Rev. Lett., 1997, {\bf 78},
2062).

\item[${\cal Q}$] {\em (Carlos V\'asquez)}: In 1983, Weinrib and Halperin
proposed a model to describe an effect of extended (correlated)
disorder on magnetic 2nd order phase transition. There, the
impurity-impurity correlation function decays for large
separations as $g(r)\sim r^{-a}$. Currently, there exist two
different predictions for the critical exponents of such a model.
What are the methods used to obtain them?

\item[${\cal A}$] Indeed, the original result of Weinrib and Halperin (
A. Weinrib,    Halperin~B.I.,  Phys. Rev. B, 1983, {\bf 27}, 417)
was obtained in the first order of the expansion in
$\varepsilon=4-d$, $\delta=4-a$. Recent two-loop estimates (V. V.
Prudnikov,  Prudnikov~P.V.,   Fedorenko~A.A.,   J. Phys. A, 1999,
{\bf 32}, L399) are due to the fixed $d$, $a$ RG technique.
Qualitative answer of both approaches is that for $a<d$ the
disorder is relevant if the correlation length critical exponent
of the system without defects obeys $\nu<2/a$.

\item[${\cal Q}$] {\em (Bertrand Berche)}: Could you comment on
logarithmic corrections to the scaling laws? Is the critical
behaviour at marginal dimensions you were speaking about (e.g. at
$m_{\mathrm{c}}^{\rm dil}\simeq1.91$ for the 3d diluted $m$-vector
magnet) governed by such corrections?

\item[${\cal A}$] Logarithmic corrections arise at upper critical
dimension $d^{\rm up}$: that is, at the {\em space} dimension,
above which the mean-field theory holds\footnote{Note added in
proof: We do not discuss here the logarithmic corrections
appearing at low dimensions as those in 2d diluted Ising model or
2d $q=4$ Potts model. For a list of systems where logarithmic
corrections appear and for the scaling relations between them see:
Kenna~R., Johnston~D.A., Janke~W., Phys. Rev. Lett., 2006,
\textbf{96}, 115701.}. For the models I was speaking about, the
upper critical dimension is four. Indeed, the divergence of an
isothermal susceptibility of the $m$-vector model at $d=4$ is
governed by the mean field exponent $\gamma=1$. However, the power
law singularity is accompanied by a logarithmic one:
 \begin{equation} \label{7.1}
 \chi^{-1} \sim|\tau|^{-1}{\ln |\tau|}^{-(m+2)/(m+8)},
 \end{equation}
as first derived in:  Larkin~A.I.,   Khmelnitskii~D.E.,  JETP, 1969,
{\bf 29}, 1123. In the RG scheme, one can see the origin of such
corrections solving the flow equation (\ref{6.1})
 at $d=4$ ($\varepsilon=0$):
  \begin{equation} \label{7.2}
  \frac{\rd
u}{\rd\ln \ell}=-u^2,
 \end{equation}
here the right-hand side is the leading term of the
$\beta$-function (\ref{6.2}). The solution of equation~(\ref{7.2})
  \begin{equation} \label{7.3}
 u=\frac{1}{|\ln \ell|} \hspace{1em} + \hspace{1em} {\rm const},
 \hspace{2em} \ell \rightarrow 0
 \end{equation}
being substituted into an appropriate expression for the
susceptibility leads to the above dependence (\ref{7.1}). Now, let
us return to the marginal dimensions $m_{\mathrm{c}}$ we were
discussing in these lectures. They are {\em field} dimensions and
we estimated them at {\em space} dimension $d=3$. As far as the
space dimension was lower than $d^{\rm up}=4$, the logarithmic
corrections do not appear. Returning back to the example given by
formulas (\ref{7.2}), (\ref{7.3}): even at $m=m_{\mathrm{c}}$, the
first power of couplings is present in the right-hand side of
functions (\ref{7.2}) for $d=3$ and the solutions will rather
behave as $u \sim \ell^{\rm const}$.

\item[${\cal Q}$] {\em (Dragi Karevski)}: Do you know other
citeria, similar to the Harris one, which predict changes in the
critical behaviour caused by different types of disorder? For
example, what happens when disorder is coupled to the order
parameter?

\item[${\cal A}$] Indeed, Harris criterion concerns the systems, where
disorder is coupled to the energy density (look for example at the
Hamiltonian (\ref{3.3}), where random variables $c_{\bf R}$ are
coupled to the product of spins). It states that critical
exponents of a disordered system do not change, if the heat
capacity of the pure system does not diverge. Later, a statement
that the correlation length critical exponent of $d$-dimensional
systems with such type of disorder should obey an inequality
$\nu>2/d$ was proven (  Chayes~J.T.,  L. Chayes,  Fisher~D.S.,  T. Spenser, Phys. Rev. Lett., 1986, {\bf 57}, 2999). For the
extended, long-range correlated disorder the generalized Harris
criterion holds (see the above mentioned paper of Weinrib and
Halperin as well as D. Boyanovsky,  Cardy~J.L.,  Phys. Rev. B, 1983,
{\bf 27}, 6971).

Y. Imry and S.-k Ma have shown (Phys. Rev. Lett., 1975, {\bf 35},
1399) that even arbitrary weak disorder coupled to the order
parameter of continuous symmetry (i.e. for $m\geqslant  2$ vector
model) destroys ferromagnetism at $d<4$. For the random-field
Ising model ($m=1$) the lower critical dimension is $d=2$. As I
already have mentioned in the lectures, the isotropically
distributed random axis destroys ferromagnetism at $d<4$. One of
the ways of showing this is to exploit the arguments similar to
those of Imry and Ma for the random-field systems (Pelcovits~R.A.,
Pytte~E., Rudnick~J., Phys. Rev. Lett., 1978, {\bf 40}, 476).

Last but not least, let me mention the Luck criterion for the
connectivity disorder, as explained in the lecture by Wolfhard
Janke.

\end{itemize}

%
%
  \label{last@page}
 \end{document}